%% file: main.tex
\documentclass[sigconf]{acmart}

\AtBeginDocument{%
  \providecommand\BibTeX{{%
    \normalfont B\kern-0.5em{\scshape i\kern-0.25em b}\kern-0.8em\TeX}}}

\setcopyright{none}
\renewcommand\footnotetextcopyrightpermission[1]{}
\copyrightyear{2027}
\acmYear{2027}

\acmConference[WWW '27]{The ACM Web Conference 2027}{2027}{}
\acmBooktitle{Proceedings of the ACM Web Conference 2027 (WWW '27)}

\usepackage{booktabs}
\usepackage{amsmath}
\usepackage{multirow}
\usepackage{graphicx}
\usepackage{placeins}
\usepackage{xcolor}
\usepackage{subcaption}
\usepackage{url}
\def\UrlBreaks{\do\/\do\-\do\_}
\makeatletter
\g@addto@macro\UrlBreaks{\do\a\do\b\do\c\do\d\do\e\do\f\do\g\do\h\do\i\do\j\do\k\do\l\do\m\do\n\do\o\do\p\do\q\do\r\do\s\do\t\do\u\do\v\do\w\do\x\do\y\do\z\do\A\do\B\do\C\do\D\do\E\do\F\do\G\do\H\do\I\do\J\do\K\do\L\do\M\do\N\do\O\do\P\do\Q\do\R\do\S\do\T\do\U\do\V\do\W\do\X\do\Y\do\Z\do\0\do\1\do\2\do\3\do\4\do\5\do\6\do\7\do\8\do\9}
\makeatother
\input{table_values}

\begin{document}

\title{ReuseGR: Offline--Online Decoupled Generative Retrieval via Adaptive Prefix Reuse}
\author{Xujing Wang}
\authornote{Equal Contribution}
\affiliation{%
  \institution{Xidian University}
  \city{Xi'an}
  \state{Shaanxi}
  \country{China}
}
\email{xjwong@stu.xidian.edu.cn}

\author{Yufeng Chen}
\authornotemark[1]
\authornote{Project leader}
\affiliation{%
  \institution{Alibaba International Digital Commerce Group}
  \city{Hangzhou}
  \state{Zhejiang}
  \country{China}
}
\email{chenyufeng.cyf@alibaba-inc.com}

\author{Boxuan Zhang}
\affiliation{%
  \institution{Alibaba International Digital Commerce Group}
  \city{Hangzhou}
  \state{Zhejiang}
  \country{China}
}
\email{boxuan.zbx@alibaba-inc.com}

\author{Chao Wei}
\affiliation{%
  \institution{Alibaba International Digital Commerce Group}
  \city{Hangzhou}
  \state{Zhejiang}
  \country{China}
}

\author{Cai Xu}
\affiliation{%
  \institution{Xidian University}
  \city{Xi'an}
  \state{Shaanxi}
  \country{China}
}
\email{cxu@xidian.edu.cn}

\author{Jie Zhao}
\affiliation{%
  \institution{Xidian University}
  \city{Xi'an}
  \state{Shaanxi}
  \country{China}
}

\author{Weiru Zhang}
\affiliation{%
  \institution{Alibaba International Digital Commerce Group}
  \city{Hangzhou}
  \state{Zhejiang}
  \country{China}
}

\author{Wei Zhao}
\affiliation{%
  \institution{Xidian University}
  \city{Xi'an}
  \state{Shaanxi}
  \country{China}
}

\author{Ziyu Guan}
\authornote{Corresponding author}
\affiliation{%
  \institution{Xidian University}
  \city{Xi'an}
  \state{Shaanxi}
  \country{China}
}
\email{zyguan@xidian.edu.cn}

\renewcommand{\shortauthors}{Wang et al.}

\begin{abstract}
Larger generative retrieval models can improve retrieval quality, but their decoding cost makes low-latency serving difficult. When different users submit the same query, generating complete Semantic IDs (SIDs) for every request repeats computation, while their interaction histories call for different results. The challenge is to reuse large-model computation across requests while keeping retrieval personalized. 

We propose \textbf{ReuseGR}, which separates query-level candidate planning from request-level personalized completion. A 7B offline planner generates SID prefix sets shared across requests with the same query. A 0.5B online model uses each request's history to select prefixes and generate the remaining tokens. ReuseCut chooses a cache depth for each query based on historical coverage, balancing computation reuse and candidate coverage.  To strengthen online completion, we transfer 7B suffix-scoring knowledge to 0.5B through path-level knowledge distillation (Path KD). BudgetCut then selects scored branches to improve final top-100 recall under a fixed branch budget. Its selection policy is initialized by on-policy distillation (OPD) and refined through group-relative policy optimization (GRPO). 

On an industrial search-advertising dataset with approximately 2M query--item records and 3.38M catalog items, ReuseGR improves query-level HR@100 by 3.83 percentage points and reduces mean online latency by 38.9\% relative to an all-online 0.5B baseline when prefixes are prepared. It achieves HR@100 comparable to an all-online 7B baseline with 57.7\% lower online latency. A 14-day user-level online A/B test against the 0.5B baseline shows relative gains of 0.93\% in click-through rate and 1.05\% in revenue per thousand requests. 
\end{abstract}



\keywords{Generative retrieval, Semantic ID, Personalized search, Adaptive prefix caching, Knowledge distillation, Sponsored search}

\maketitle

\input{sections/01_introduction}
\input{sections/02_related}
\input{sections/03_preliminary}
\input{sections/04_method}
\input{sections/05_experiments}

\input{sections/06_discussion}
\input{sections/07_conclusion}

\label{sec:main-end}
\clearpage
\bibliographystyle{ACM-Reference-Format}
\bibliography{references}

\input{sections/08_appendix}

\end{document}

%% file: table_values.tex
\newcommand*{\TblVal}[1]{\ifcsname tblval@#1\endcsname\csname tblval@#1\endcsname\else\PackageError{ReuseGR}{Unknown table value #1}{}\fi}
\expandafter\def\csname tblval@adaptive.d\endcsname{1.72}
\expandafter\def\csname tblval@adaptive.steps\endcsname{2.28}
\expandafter\def\csname tblval@cohort.head.queries\endcsname{4,120}
\expandafter\def\csname tblval@cohort.test.queries\endcsname{44,225}
\expandafter\def\csname tblval@cohort.torso.queries\endcsname{5,156}
\expandafter\def\csname tblval@cohort.unseen.queries\endcsname{34,949}
\expandafter\def\csname tblval@diagnostic.layer1.mode\endcsname{5.70}
\expandafter\def\csname tblval@diagnostic.layer1.mode-gain\endcsname{+59.24}
\expandafter\def\csname tblval@diagnostic.layer1.random\endcsname{3.12}
\expandafter\def\csname tblval@diagnostic.layer1.scale-gain\endcsname{+1.79}
\expandafter\def\csname tblval@diagnostic.layer1.small\endcsname{63.15}
\expandafter\def\csname tblval@diagnostic.layer1.teacher\endcsname{64.93}
\expandafter\def\csname tblval@diagnostic.layer1.vocab\endcsname{32.00}
\expandafter\def\csname tblval@diagnostic.layer2.mode\endcsname{13.60}
\expandafter\def\csname tblval@diagnostic.layer2.mode-gain\endcsname{+29.75}
\expandafter\def\csname tblval@diagnostic.layer2.random\endcsname{3.19}
\expandafter\def\csname tblval@diagnostic.layer2.scale-gain\endcsname{+1.86}
\expandafter\def\csname tblval@diagnostic.layer2.small\endcsname{41.49}
\expandafter\def\csname tblval@diagnostic.layer2.teacher\endcsname{43.35}
\expandafter\def\csname tblval@diagnostic.layer2.vocab\endcsname{31.36}
\expandafter\def\csname tblval@diagnostic.layer3.mode\endcsname{24.24}
\expandafter\def\csname tblval@diagnostic.layer3.mode-gain\endcsname{+19.03}
\expandafter\def\csname tblval@diagnostic.layer3.random\endcsname{3.35}
\expandafter\def\csname tblval@diagnostic.layer3.scale-gain\endcsname{+1.69}
\expandafter\def\csname tblval@diagnostic.layer3.small\endcsname{41.58}
\expandafter\def\csname tblval@diagnostic.layer3.teacher\endcsname{43.27}
\expandafter\def\csname tblval@diagnostic.layer3.vocab\endcsname{29.91}
\expandafter\def\csname tblval@diagnostic.layer4.mode\endcsname{33.22}
\expandafter\def\csname tblval@diagnostic.layer4.mode-gain\endcsname{+11.50}
\expandafter\def\csname tblval@diagnostic.layer4.random\endcsname{5.18}
\expandafter\def\csname tblval@diagnostic.layer4.scale-gain\endcsname{+2.53}
\expandafter\def\csname tblval@diagnostic.layer4.small\endcsname{42.19}
\expandafter\def\csname tblval@diagnostic.layer4.teacher\endcsname{44.72}
\expandafter\def\csname tblval@diagnostic.layer4.vocab\endcsname{21.99}
\expandafter\def\csname tblval@drop.l1.seen.n\endcsname{18,748}
\expandafter\def\csname tblval@drop.l1.seen.share\endcsname{4.9031}
\expandafter\def\csname tblval@drop.l1.train.n\endcsname{10,011}
\expandafter\def\csname tblval@drop.l1.train.share\endcsname{0.9794}
\expandafter\def\csname tblval@drop.l2.seen.n\endcsname{48,238}
\expandafter\def\csname tblval@drop.l2.seen.share\endcsname{12.6154}
\expandafter\def\csname tblval@drop.l2.train.n\endcsname{98,158}
\expandafter\def\csname tblval@drop.l2.train.share\endcsname{9.6031}
\expandafter\def\csname tblval@drop.l3.seen.n\endcsname{75,784}
\expandafter\def\csname tblval@drop.l3.seen.share\endcsname{19.8193}
\expandafter\def\csname tblval@drop.l3.train.n\endcsname{364,496}
\expandafter\def\csname tblval@drop.l3.train.share\endcsname{35.6597}
\expandafter\def\csname tblval@drop.l4.seen.n\endcsname{92,667}
\expandafter\def\csname tblval@drop.l4.seen.share\endcsname{24.2346}
\expandafter\def\csname tblval@drop.l4.train.n\endcsname{295,078}
\expandafter\def\csname tblval@drop.l4.train.share\endcsname{28.8684}
\expandafter\def\csname tblval@drop.none.seen.n\endcsname{140,915}
\expandafter\def\csname tblval@drop.none.seen.share\endcsname{36.8527}
\expandafter\def\csname tblval@drop.none.train.n\endcsname{223,861}
\expandafter\def\csname tblval@drop.none.train.share\endcsname{21.9010}
\expandafter\def\csname tblval@drop.total.seen.n\endcsname{382,374}
\expandafter\def\csname tblval@drop.total.seen.share\endcsname{100.0000}
\expandafter\def\csname tblval@drop.total.train.n\endcsname{1,022,150}
\expandafter\def\csname tblval@drop.total.train.share\endcsname{100.0000}
\expandafter\def\csname tblval@drop.unknown.seen.n\endcsname{6,022}
\expandafter\def\csname tblval@drop.unknown.seen.share\endcsname{1.5749}
\expandafter\def\csname tblval@drop.unknown.train.n\endcsname{30,546}
\expandafter\def\csname tblval@drop.unknown.train.share\endcsname{2.9884}
\expandafter\def\csname tblval@fixed2.d\endcsname{2.00}
\expandafter\def\csname tblval@fixed2.steps\endcsname{2.00}
\expandafter\def\csname tblval@frontier-adaptive.ch\endcsname{97.84}
\expandafter\def\csname tblval@frontier-adaptive.cr\endcsname{78.62}
\expandafter\def\csname tblval@frontier-fixed2.ch\endcsname{96.69}
\expandafter\def\csname tblval@frontier-fixed2.cr\endcsname{74.18}
\expandafter\def\csname tblval@frontier-permuted.ch\endcsname{95.76}
\expandafter\def\csname tblval@item.dssm.hr100\endcsname{69.84}
\expandafter\def\csname tblval@item.dssm.hr500\endcsname{75.33}
\expandafter\def\csname tblval@item.dssm.rec100\endcsname{26.12}
\expandafter\def\csname tblval@item.dssm.rec500\endcsname{38.40}
\expandafter\def\csname tblval@item.fixed2-b15.expand\endcsname{0.35}
\expandafter\def\csname tblval@item.fixed2-b15.hr100\endcsname{76.85}
\expandafter\def\csname tblval@item.fixed2-b15.hr500\endcsname{81.68}
\expandafter\def\csname tblval@item.fixed2-b15.rec100\endcsname{34.28}
\expandafter\def\csname tblval@item.fixed2-b15.rec500\endcsname{44.15}
\expandafter\def\csname tblval@item.full.expand\endcsname{0.35}
\expandafter\def\csname tblval@item.full.hr100\endcsname{80.64}
\expandafter\def\csname tblval@item.full.hr500\endcsname{85.36}
\expandafter\def\csname tblval@item.full.rec100\endcsname{36.14}
\expandafter\def\csname tblval@item.full.rec500\endcsname{44.68}
\expandafter\def\csname tblval@item.onerec-think.expand\endcsname{0.35}
\expandafter\def\csname tblval@item.onerec-think.hr100\endcsname{75.96}
\expandafter\def\csname tblval@item.onerec-think.hr500\endcsname{81.08}
\expandafter\def\csname tblval@item.onerec-think.rec100\endcsname{33.40}
\expandafter\def\csname tblval@item.onerec-think.rec500\endcsname{43.35}
\expandafter\def\csname tblval@item.small.expand\endcsname{0.35}
\expandafter\def\csname tblval@item.small.hr100\endcsname{76.42}
\expandafter\def\csname tblval@item.small.hr500\endcsname{81.54}
\expandafter\def\csname tblval@item.small.rec100\endcsname{33.85}
\expandafter\def\csname tblval@item.small.rec500\endcsname{43.92}
\expandafter\def\csname tblval@item.smartgr.expand\endcsname{0.35}
\expandafter\def\csname tblval@item.smartgr.hr100\endcsname{78.24}
\expandafter\def\csname tblval@item.smartgr.hr500\endcsname{83.12}
\expandafter\def\csname tblval@item.smartgr.rec100\endcsname{34.85}
\expandafter\def\csname tblval@item.smartgr.rec500\endcsname{44.32}
\expandafter\def\csname tblval@item.t5.expand\endcsname{0.35}
\expandafter\def\csname tblval@item.t5.hr100\endcsname{72.60}
\expandafter\def\csname tblval@item.t5.hr500\endcsname{77.82}
\expandafter\def\csname tblval@item.t5.rec100\endcsname{28.94}
\expandafter\def\csname tblval@item.t5.rec500\endcsname{37.95}
\expandafter\def\csname tblval@item.teacher.expand\endcsname{0.35}
\expandafter\def\csname tblval@item.teacher.hr100\endcsname{80.58}
\expandafter\def\csname tblval@item.teacher.hr500\endcsname{85.28}
\expandafter\def\csname tblval@item.teacher.rec100\endcsname{37.42}
\expandafter\def\csname tblval@item.teacher.rec500\endcsname{49.10}
\expandafter\def\csname tblval@probe.layer1.gain\endcsname{+61.18}
\expandafter\def\csname tblval@probe.layer1.support\endcsname{48.60}
\expandafter\def\csname tblval@probe.layer2.gain\endcsname{+25.95}
\expandafter\def\csname tblval@probe.layer2.support\endcsname{36.15}
\expandafter\def\csname tblval@probe.layer3.ci.high\endcsname{6.78}
\expandafter\def\csname tblval@probe.layer3.ci.low\endcsname{0.25}
\expandafter\def\csname tblval@probe.layer3.gain\endcsname{+3.51}
\expandafter\def\csname tblval@probe.layer3.support\endcsname{19.00}
\expandafter\def\csname tblval@probe.layer4.ci.high\endcsname{3.20}
\expandafter\def\csname tblval@probe.layer4.ci.low\endcsname{-6.01}
\expandafter\def\csname tblval@probe.layer4.gain\endcsname{-1.57}
\expandafter\def\csname tblval@probe.layer4.support\endcsname{9.15}
\expandafter\def\csname tblval@q.adaptive-kd.hr1\endcsname{36.99}
\expandafter\def\csname tblval@q.adaptive-kd.hr10\endcsname{66.36}
\expandafter\def\csname tblval@q.adaptive-kd.hr100\endcsname{83.32}
\expandafter\def\csname tblval@q.adaptive-kd.mean\endcsname{76.09}
\expandafter\def\csname tblval@q.adaptive-kd.rec1\endcsname{6.22}
\expandafter\def\csname tblval@q.adaptive-kd.rec10\endcsname{23.26}
\expandafter\def\csname tblval@q.adaptive-kd.rec100\endcsname{44.76}
\expandafter\def\csname tblval@q.adaptive-sft.hr1\endcsname{36.62}
\expandafter\def\csname tblval@q.adaptive-sft.hr10\endcsname{65.98}
\expandafter\def\csname tblval@q.adaptive-sft.hr100\endcsname{82.64}
\expandafter\def\csname tblval@q.adaptive-sft.mean\endcsname{76.09}
\expandafter\def\csname tblval@q.adaptive-sft.rec1\endcsname{6.11}
\expandafter\def\csname tblval@q.adaptive-sft.rec10\endcsname{22.91}
\expandafter\def\csname tblval@q.adaptive-sft.rec100\endcsname{44.38}
\expandafter\def\csname tblval@q.dssm.hr1\endcsname{15.21}
\expandafter\def\csname tblval@q.dssm.hr10\endcsname{48.91}
\expandafter\def\csname tblval@q.dssm.hr100\endcsname{75.33}
\expandafter\def\csname tblval@q.dssm.mean\endcsname{73.32}
\expandafter\def\csname tblval@q.dssm.rec1\endcsname{2.22}
\expandafter\def\csname tblval@q.dssm.rec10\endcsname{13.05}
\expandafter\def\csname tblval@q.dssm.rec100\endcsname{38.40}
\expandafter\def\csname tblval@q.dssm.speedup\endcsname{2.30}
\expandafter\def\csname tblval@q.fixed1.ch\endcsname{100.00}
\expandafter\def\csname tblval@q.fixed1.cr\endcsname{100.00}
\expandafter\def\csname tblval@q.fixed1.hr1\endcsname{36.08}
\expandafter\def\csname tblval@q.fixed1.hr10\endcsname{65.42}
\expandafter\def\csname tblval@q.fixed1.hr100\endcsname{82.01}
\expandafter\def\csname tblval@q.fixed1.mean\endcsname{93.16}
\expandafter\def\csname tblval@q.fixed1.rec1\endcsname{5.81}
\expandafter\def\csname tblval@q.fixed1.rec10\endcsname{22.21}
\expandafter\def\csname tblval@q.fixed1.rec100\endcsname{45.60}
\expandafter\def\csname tblval@q.fixed2-b1.hr1\endcsname{36.35}
\expandafter\def\csname tblval@q.fixed2-b1.hr10\endcsname{65.71}
\expandafter\def\csname tblval@q.fixed2-b1.hr100\endcsname{82.04}
\expandafter\def\csname tblval@q.fixed2-b1.mean\endcsname{69.33}
\expandafter\def\csname tblval@q.fixed2-b1.rec1\endcsname{5.99}
\expandafter\def\csname tblval@q.fixed2-b1.rec10\endcsname{22.58}
\expandafter\def\csname tblval@q.fixed2-b1.rec100\endcsname{45.65}
\expandafter\def\csname tblval@q.fixed2-b15.hr1\endcsname{36.48}
\expandafter\def\csname tblval@q.fixed2-b15.hr10\endcsname{65.84}
\expandafter\def\csname tblval@q.fixed2-b15.hr100\endcsname{82.11}
\expandafter\def\csname tblval@q.fixed2-b15.mean\endcsname{69.33}
\expandafter\def\csname tblval@q.fixed2-b15.rec1\endcsname{6.05}
\expandafter\def\csname tblval@q.fixed2-b15.rec10\endcsname{22.70}
\expandafter\def\csname tblval@q.fixed2-b15.rec100\endcsname{45.72}
\expandafter\def\csname tblval@q.fixed2-full-retrained.hr1\endcsname{37.65}
\expandafter\def\csname tblval@q.fixed2-full-retrained.hr10\endcsname{67.18}
\expandafter\def\csname tblval@q.fixed2-full-retrained.hr100\endcsname{84.15}
\expandafter\def\csname tblval@q.fixed2-full-retrained.mean\endcsname{67.25}
\expandafter\def\csname tblval@q.fixed2-full-retrained.rec100\endcsname{45.40}
\expandafter\def\csname tblval@q.fixed2-full-switch.hr1\endcsname{37.42}
\expandafter\def\csname tblval@q.fixed2-full-switch.hr10\endcsname{66.85}
\expandafter\def\csname tblval@q.fixed2-full-switch.hr100\endcsname{83.76}
\expandafter\def\csname tblval@q.fixed2-full-switch.mean\endcsname{67.25}
\expandafter\def\csname tblval@q.fixed2-full-switch.rec100\endcsname{45.12}
\expandafter\def\csname tblval@q.fixed3.ch\endcsname{89.14}
\expandafter\def\csname tblval@q.fixed3.cr\endcsname{56.40}
\expandafter\def\csname tblval@q.fixed3.hr1\endcsname{35.94}
\expandafter\def\csname tblval@q.fixed3.hr10\endcsname{64.38}
\expandafter\def\csname tblval@q.fixed3.hr100\endcsname{80.12}
\expandafter\def\csname tblval@q.fixed3.mean\endcsname{45.50}
\expandafter\def\csname tblval@q.fixed3.rec1\endcsname{5.89}
\expandafter\def\csname tblval@q.fixed3.rec10\endcsname{21.78}
\expandafter\def\csname tblval@q.fixed3.rec100\endcsname{42.45}
\expandafter\def\csname tblval@q.full.hr1\endcsname{38.74}
\expandafter\def\csname tblval@q.full.hr10\endcsname{68.53}
\expandafter\def\csname tblval@q.full.hr100\endcsname{85.82}
\expandafter\def\csname tblval@q.full.mean\endcsname{71.45}
\expandafter\def\csname tblval@q.full.rec1\endcsname{6.52}
\expandafter\def\csname tblval@q.full.rec10\endcsname{24.16}
\expandafter\def\csname tblval@q.full.rec100\endcsname{46.05}
\expandafter\def\csname tblval@q.full.speedup\endcsname{2.36}
\expandafter\def\csname tblval@q.heuristic.hr1\endcsname{36.84}
\expandafter\def\csname tblval@q.heuristic.hr10\endcsname{66.15}
\expandafter\def\csname tblval@q.heuristic.hr100\endcsname{82.91}
\expandafter\def\csname tblval@q.heuristic.mean\endcsname{71.03}
\expandafter\def\csname tblval@q.heuristic.rec1\endcsname{6.18}
\expandafter\def\csname tblval@q.heuristic.rec10\endcsname{23.08}
\expandafter\def\csname tblval@q.heuristic.rec100\endcsname{44.52}
\expandafter\def\csname tblval@q.onerec-think.hr1\endcsname{35.45}
\expandafter\def\csname tblval@q.onerec-think.hr10\endcsname{64.92}
\expandafter\def\csname tblval@q.onerec-think.hr100\endcsname{81.48}
\expandafter\def\csname tblval@q.onerec-think.mean\endcsname{74.20}
\expandafter\def\csname tblval@q.onerec-think.rec1\endcsname{5.72}
\expandafter\def\csname tblval@q.onerec-think.rec10\endcsname{21.75}
\expandafter\def\csname tblval@q.onerec-think.rec100\endcsname{44.92}
\expandafter\def\csname tblval@q.onerec-think.speedup\endcsname{2.28}
\expandafter\def\csname tblval@q.opd.hr1\endcsname{37.83}
\expandafter\def\csname tblval@q.opd.hr10\endcsname{67.34}
\expandafter\def\csname tblval@q.opd.hr100\endcsname{84.20}
\expandafter\def\csname tblval@q.opd.mean\endcsname{71.45}
\expandafter\def\csname tblval@q.opd.rec1\endcsname{6.36}
\expandafter\def\csname tblval@q.opd.rec10\endcsname{23.89}
\expandafter\def\csname tblval@q.opd.rec100\endcsname{45.32}
\expandafter\def\csname tblval@q.permuted-full.hr1\endcsname{36.65}
\expandafter\def\csname tblval@q.permuted-full.hr10\endcsname{66.10}
\expandafter\def\csname tblval@q.permuted-full.hr100\endcsname{82.65}
\expandafter\def\csname tblval@q.permuted-full.mean\endcsname{73.10}
\expandafter\def\csname tblval@q.permuted-full.rec100\endcsname{43.85}
\expandafter\def\csname tblval@q.permuted-sft.hr1\endcsname{36.05}
\expandafter\def\csname tblval@q.permuted-sft.hr10\endcsname{65.13}
\expandafter\def\csname tblval@q.permuted-sft.hr100\endcsname{81.44}
\expandafter\def\csname tblval@q.permuted-sft.mean\endcsname{76.10}
\expandafter\def\csname tblval@q.permuted-sft.rec100\endcsname{44.63}
\expandafter\def\csname tblval@q.popularity.hr1\endcsname{31.45}
\expandafter\def\csname tblval@q.popularity.hr10\endcsname{58.92}
\expandafter\def\csname tblval@q.popularity.hr100\endcsname{76.48}
\expandafter\def\csname tblval@q.popularity.mean\endcsname{69.33}
\expandafter\def\csname tblval@q.popularity.rec1\endcsname{4.92}
\expandafter\def\csname tblval@q.popularity.rec10\endcsname{19.15}
\expandafter\def\csname tblval@q.popularity.rec100\endcsname{39.84}
\expandafter\def\csname tblval@q.seen.ch\endcsname{99.12}
\expandafter\def\csname tblval@q.seen.cr\endcsname{86.45}
\expandafter\def\csname tblval@q.seen.hr100\endcsname{94.68}
\expandafter\def\csname tblval@q.seen.rec100\endcsname{54.12}
\expandafter\def\csname tblval@q.small-planner-fixed.hr1\endcsname{35.58}
\expandafter\def\csname tblval@q.small-planner-fixed.hr10\endcsname{64.96}
\expandafter\def\csname tblval@q.small-planner-fixed.hr100\endcsname{81.82}
\expandafter\def\csname tblval@q.small-planner-fixed.mean\endcsname{69.33}
\expandafter\def\csname tblval@q.small-planner-fixed.rec1\endcsname{5.57}
\expandafter\def\csname tblval@q.small-planner-fixed.rec10\endcsname{21.74}
\expandafter\def\csname tblval@q.small-planner-fixed.rec100\endcsname{45.38}
\expandafter\def\csname tblval@q.small-planner-full.hr1\endcsname{36.85}
\expandafter\def\csname tblval@q.small-planner-full.hr10\endcsname{66.42}
\expandafter\def\csname tblval@q.small-planner-full.hr100\endcsname{83.46}
\expandafter\def\csname tblval@q.small-planner-full.mean\endcsname{71.45}
\expandafter\def\csname tblval@q.small-planner-full.rec1\endcsname{6.08}
\expandafter\def\csname tblval@q.small-planner-full.rec10\endcsname{22.84}
\expandafter\def\csname tblval@q.small-planner-full.rec100\endcsname{45.69}
\expandafter\def\csname tblval@q.small.ch\endcsname{100.00}
\expandafter\def\csname tblval@q.small.cr\endcsname{100.00}
\expandafter\def\csname tblval@q.small.hr1\endcsname{35.72}
\expandafter\def\csname tblval@q.small.hr10\endcsname{65.13}
\expandafter\def\csname tblval@q.small.hr100\endcsname{81.99}
\expandafter\def\csname tblval@q.small.mean\endcsname{117.00}
\expandafter\def\csname tblval@q.small.rec1\endcsname{5.61}
\expandafter\def\csname tblval@q.small.rec10\endcsname{21.83}
\expandafter\def\csname tblval@q.small.rec100\endcsname{45.53}
\expandafter\def\csname tblval@q.small.speedup\endcsname{1.44}
\expandafter\def\csname tblval@q.smartgr.hr1\endcsname{36.82}
\expandafter\def\csname tblval@q.smartgr.hr10\endcsname{66.75}
\expandafter\def\csname tblval@q.smartgr.hr100\endcsname{83.56}
\expandafter\def\csname tblval@q.smartgr.mean\endcsname{114.20}
\expandafter\def\csname tblval@q.smartgr.rec1\endcsname{5.92}
\expandafter\def\csname tblval@q.smartgr.rec10\endcsname{22.84}
\expandafter\def\csname tblval@q.smartgr.rec100\endcsname{46.90}
\expandafter\def\csname tblval@q.smartgr.speedup\endcsname{1.48}
\expandafter\def\csname tblval@q.t5.hr1\endcsname{31.98}
\expandafter\def\csname tblval@q.t5.hr10\endcsname{59.95}
\expandafter\def\csname tblval@q.t5.hr100\endcsname{78.28}
\expandafter\def\csname tblval@q.t5.mean\endcsname{211.62}
\expandafter\def\csname tblval@q.t5.rec1\endcsname{5.50}
\expandafter\def\csname tblval@q.t5.rec10\endcsname{18.87}
\expandafter\def\csname tblval@q.t5.rec100\endcsname{39.38}
\expandafter\def\csname tblval@q.t5.speedup\endcsname{0.80}
\expandafter\def\csname tblval@q.teacher.hr1\endcsname{39.62}
\expandafter\def\csname tblval@q.teacher.hr10\endcsname{69.94}
\expandafter\def\csname tblval@q.teacher.hr100\endcsname{85.71}
\expandafter\def\csname tblval@q.teacher.mean\endcsname{168.88}
\expandafter\def\csname tblval@q.teacher.rec1\endcsname{6.43}
\expandafter\def\csname tblval@q.teacher.rec10\endcsname{24.79}
\expandafter\def\csname tblval@q.teacher.rec100\endcsname{50.96}
\expandafter\def\csname tblval@q.teacher.speedup\endcsname{1.00}
\expandafter\def\csname tblval@q.unseen.ch\endcsname{97.50}
\expandafter\def\csname tblval@q.unseen.cr\endcsname{76.54}
\expandafter\def\csname tblval@q.unseen.hr100\endcsname{83.47}
\expandafter\def\csname tblval@q.unseen.rec100\endcsname{43.91}
\expandafter\def\csname tblval@row.cache100.bytes\endcsname{400}
\expandafter\def\csname tblval@row.cache100.hr1\endcsname{37.15}
\expandafter\def\csname tblval@row.cache100.hr100\endcsname{84.18}
\expandafter\def\csname tblval@row.cache100.mean\endcsname{0.18}
\expandafter\def\csname tblval@row.canonical.bytes\endcsname{256}
\expandafter\def\csname tblval@row.canonical.hr1\endcsname{40.48}
\expandafter\def\csname tblval@row.canonical.hr100\endcsname{88.34}
\expandafter\def\csname tblval@row.canonical.mean\endcsname{71.45}
\expandafter\def\csname tblval@row.empty.bytes\endcsname{256}
\expandafter\def\csname tblval@row.empty.hr1\endcsname{38.85}
\expandafter\def\csname tblval@row.empty.hr100\endcsname{87.88}
\expandafter\def\csname tblval@row.empty.mean\endcsname{71.45}
\expandafter\def\csname tblval@row.real.bytes\endcsname{256}
\expandafter\def\csname tblval@row.real.hr1\endcsname{41.22}
\expandafter\def\csname tblval@row.real.hr100\endcsname{88.95}
\expandafter\def\csname tblval@row.real.mean\endcsname{71.45}
\expandafter\def\csname tblval@row.rerank100.bytes\endcsname{400}
\expandafter\def\csname tblval@row.rerank100.hr1\endcsname{39.52}
\expandafter\def\csname tblval@row.rerank100.hr100\endcsname{84.18}
\expandafter\def\csname tblval@row.rerank100.mean\endcsname{13.00}
\expandafter\def\csname tblval@row.rerank1000.bytes\endcsname{4,000}
\expandafter\def\csname tblval@row.rerank1000.hr1\endcsname{40.55}
\expandafter\def\csname tblval@row.rerank1000.hr100\endcsname{87.62}
\expandafter\def\csname tblval@row.rerank1000.mean\endcsname{46.40}
\expandafter\def\csname tblval@row.rerank500.bytes\endcsname{2,000}
\expandafter\def\csname tblval@row.rerank500.hr1\endcsname{40.18}
\expandafter\def\csname tblval@row.rerank500.hr100\endcsname{86.85}
\expandafter\def\csname tblval@row.rerank500.mean\endcsname{28.70}
\expandafter\def\csname tblval@row.shuffled.bytes\endcsname{256}
\expandafter\def\csname tblval@row.shuffled.hr1\endcsname{38.64}
\expandafter\def\csname tblval@row.shuffled.hr100\endcsname{87.52}
\expandafter\def\csname tblval@row.shuffled.mean\endcsname{71.45}
\expandafter\def\csname tblval@serving.backfill.hr100\endcsname{84.23}
\expandafter\def\csname tblval@serving.backfill.mean\endcsname{90.40}
\expandafter\def\csname tblval@serving.backfill.pshare\endcsname{58.40}
\expandafter\def\csname tblval@serving.backfill.qshare\endcsname{42.15}
\expandafter\def\csname tblval@serving.cold.pshare\endcsname{0.00}
\expandafter\def\csname tblval@serving.cold.qshare\endcsname{0.00}
\expandafter\def\csname tblval@serving.prepared.pshare\endcsname{100.00}
\expandafter\def\csname tblval@serving.prepared.qshare\endcsname{100.00}
\expandafter\def\csname tblval@serving.rolling.hr100\endcsname{85.21}
\expandafter\def\csname tblval@serving.rolling.mean\endcsname{78.65}
\expandafter\def\csname tblval@serving.rolling.pshare\endcsname{84.20}
\expandafter\def\csname tblval@serving.rolling.qshare\endcsname{68.50}
\expandafter\def\csname tblval@serving.static-mix.hr100\endcsname{83.40}
\expandafter\def\csname tblval@serving.static-mix.mean\endcsname{100.21}
\expandafter\def\csname tblval@serving.static-mix.pshare\endcsname{36.87}
\expandafter\def\csname tblval@serving.static-mix.qshare\endcsname{20.97}
\expandafter\def\csname tblval@serving.static-q.hr100\endcsname{82.38}
\expandafter\def\csname tblval@serving.static-q.mean\endcsname{100.21}
\expandafter\def\csname tblval@serving.static-q.pshare\endcsname{36.87}
\expandafter\def\csname tblval@serving.static-q.qshare\endcsname{20.97}
\expandafter\def\csname tblval@strata.d.head\endcsname{2.68}
\expandafter\def\csname tblval@strata.d.torso\endcsname{2.14}
\expandafter\def\csname tblval@strata.d.unseen\endcsname{1.54}
\expandafter\def\csname tblval@strata.fixed2-b15.head\endcsname{87.90}
\expandafter\def\csname tblval@strata.fixed2-b15.torso\endcsname{84.82}
\expandafter\def\csname tblval@strata.fixed2-b15.unseen\endcsname{60.44}
\expandafter\def\csname tblval@strata.full.head\endcsname{91.24}
\expandafter\def\csname tblval@strata.full.torso\endcsname{87.82}
\expandafter\def\csname tblval@strata.full.unseen\endcsname{63.01}
\expandafter\def\csname tblval@strata.mean.head\endcsname{55.24}
\expandafter\def\csname tblval@strata.mean.torso\endcsname{68.14}
\expandafter\def\csname tblval@strata.mean.unseen\endcsname{73.85}
\expandafter\def\csname tblval@strata.share.all\endcsname{100.00}
\expandafter\def\csname tblval@strata.share.head\endcsname{9.32}
\expandafter\def\csname tblval@strata.share.torso\endcsname{11.66}
\expandafter\def\csname tblval@strata.share.unseen\endcsname{79.03}
\expandafter\def\csname tblval@strata.small.head\endcsname{84.15}
\expandafter\def\csname tblval@strata.small.torso\endcsname{81.60}
\expandafter\def\csname tblval@strata.small.unseen\endcsname{60.46}
\expandafter\def\csname tblval@strata.teacher.head\endcsname{91.85}
\expandafter\def\csname tblval@strata.teacher.torso\endcsname{88.90}
\expandafter\def\csname tblval@strata.teacher.unseen\endcsname{64.56}

%% file: sections/01_introduction.tex
\section{Introduction}
\label{sec:intro}

\begin{figure}[!t]
\centering
\begin{subfigure}[t]{0.49\columnwidth}
  \centering
  \includegraphics[width=\linewidth,height=4.1cm,keepaspectratio]{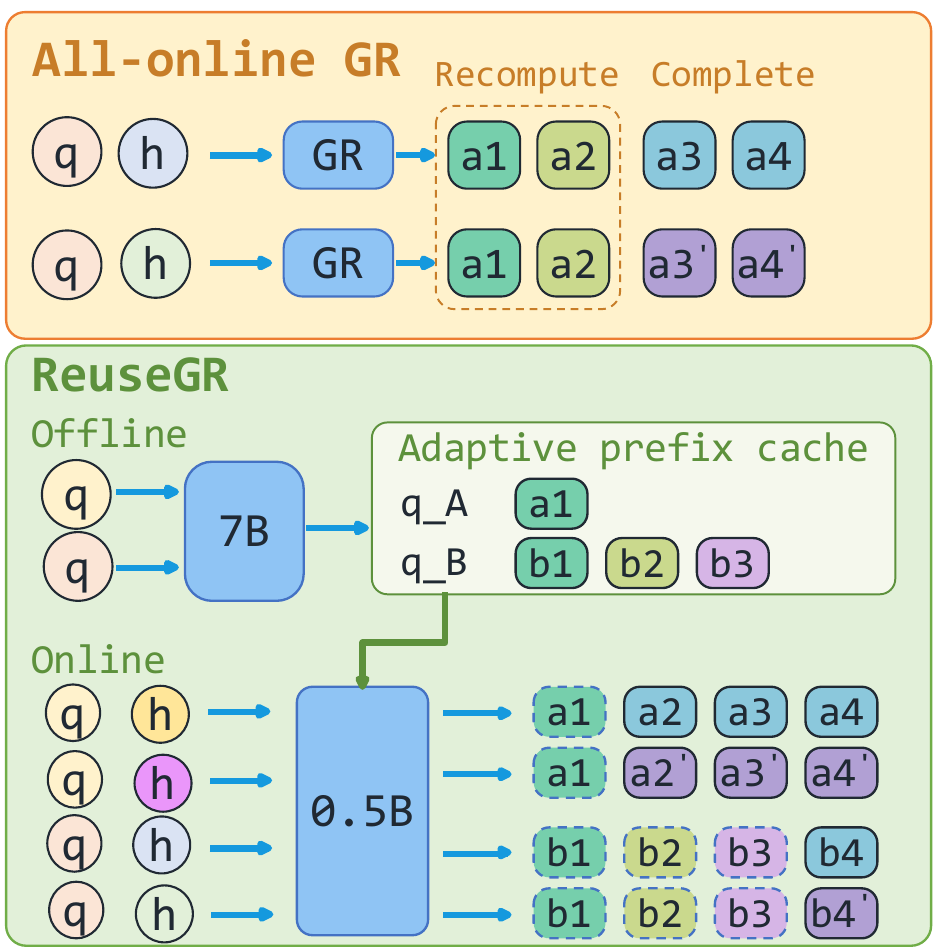}
  \caption{Prefix reuse}
  \label{fig:teaser-a}
\end{subfigure}\hfill
\begin{subfigure}[t]{0.49\columnwidth}
  \centering
  \includegraphics[width=\linewidth,height=4.1cm,keepaspectratio]{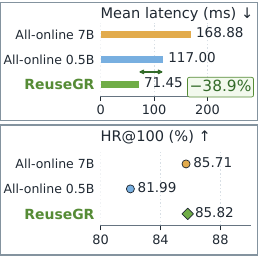}
  \caption{Efficiency and quality}
  \label{fig:teaser-b}
\end{subfigure}
\caption{Query-prefix reuse in ReuseGR. (a) Queries reuse different prefix
lengths (one and three tokens shown, dashed) and use their histories to
complete suffixes. Cache depths range from zero to three. (b) Complete-SID HR@100 and
mean online latency on AE-PV (44,225 queries, one A100, batch size 1), with
cached prefixes available for ReuseGR.}
\Description{Left: full online retrieval repeats prefix decoding for requests
with the same query and different histories. ReuseGR caches one token for
query q\_A and three for q\_B. Requests for each query share these prefixes
(dashed outlines) and use their histories for 0.5B suffix completion. 
Right: latency
and HR at 100 compare 7B and 0.5B end-to-end retrieval with ReuseGR.}
\label{fig:teaser}
\end{figure}

Search-advertising systems select candidates from million-item catalogs
for downstream ranking, making candidate quality central to user experience
and advertising value. Traditional retrieval matches queries
against indexed item representations~\cite{huang2013dssm,karpukhin2020dpr}. 
Generative retrieval learns to predict candidate identifiers directly, 
allowing semantic knowledge in the model to guide candidate
selection~\cite{tay2022transformer,wang2022nci}. 
Semantic IDs (SIDs) represent items as short token sequences whose prefixes
group related items~\cite{rajput2023recommender}. This structure lets a
generator select candidates through successive token predictions. 
As illustrated in the upper half of Figure~\ref{fig:teaser-a}, the model
reads a query and user history, then decodes complete SIDs one token at a time. 

Larger generators improve retrieval quality but increase online decoding
cost. Existing methods address this cost in several ways: distillation
teaches a smaller decoder or lightweight prediction heads to imitate a
large teacher, reducing computation per prediction~\cite{smartgr2024,sidmlp2024}; 
parallel prediction produces SID tokens in the same model call instead
of waiting for each preceding token, reducing sequential computation~\cite{hou2025rpg}; speculative decoding lets a small model draft
candidate sequences for batched verification by a large
model~\cite{lin2025atspeed}; prompt caching reuses attention states for
shared input blocks, such as history and item representations, reducing
repeated context processing~\cite{zhao2026rcllm}.
These approaches lower the cost of online prediction, while candidate
SIDs are still generated for each search. Repeated queries offer another
source of savings: sharing large-model planning across users. 

We ask whether part of this planning can be performed offline and reused
by users who submit the same query. The upper half of
Figure~\ref{fig:teaser-a} makes the opportunity concrete: two histories
lead to different complete SIDs, yet their shared initial tokens are
recomputed for both searches. Our idea is to cache query-dependent SID
prefixes and let the current history guide their completion online. 
Each prefix preserves a subtree with multiple continuations, so the
online model can generate candidates beyond a precomputed list of
complete SIDs. This would share expensive planning while keeping final
candidate selection responsive to the user. 

Our click-log analysis examines the feasibility of this idea and the
choice of prefix length. When predicting the next SID token at matched
preceding prefixes, query information contributes much more at the first
two layers than at later layers (Figure~\ref{fig:keyprobe}). This supports
sharing coarse candidate structure across users. The suitable sharing
boundary also varies by query: model-generated prefix sets at
query-specific depths retain \TblVal{frontier-adaptive.cr}\% of target
SIDs on average, versus \TblVal{frontier-fixed2.cr}\% at a uniform
two-token depth under the same prefix-count limit. These are reachable
target fractions before online completion. A deeper cut saves decoding
steps but narrows the available subtrees; a shallower cut leaves more
choices for history-conditioned generation. The design must therefore
identify reusable prefixes and select how far each query can reuse them. 

We implement this design as \textbf{ReuseGR}. In the lower half of
Figure~\ref{fig:teaser-a}, query $q_A$ reuses one token and $q_B$ reuses
three; users searching for either query share its prefixes while their
histories guide different suffixes. ReuseCut predicts these depths from
query text using historical target-coverage labels. A 7B offline planner
produces the shared prefix sets, and a 0.5B online model completes them
using the current query and history. To guide the small model's suffix
scoring with large-model knowledge, we apply path-level knowledge
distillation (Path KD) from the 7B teacher. After scoring, branches are
merged to form the final retrieval results, so branch selection affects
which relevant SIDs enter the top-100 ranking. We develop BudgetCut to
learn this selection under a fixed branch budget. On-policy distillation
(OPD) initializes its policy from marginal recall gains, and group-relative
policy optimization (GRPO) refines the policy using the recall of complete
selected branch sets. 

On AE-PV, an industrial search-advertising dataset with 3.38M items and
44,225 Test queries, ReuseGR achieves \TblVal{q.full.hr100}\% query-level
HR@100 and \TblVal{q.full.rec100}\% Recall@100 at
\TblVal{q.full.mean}\,ms mean online latency with prepared prefixes. 
It gains 3.83 HR@100 points and reduces latency by 38.9\% over an
all-online 0.5B baseline, with HR@100 comparable to all-online 7B at
57.7\% lower latency (Figure~\ref{fig:teaser-b}). Matching depths to
queries improves HR@100 by 3.17 points over permuted assignments with
the same depth distribution. With actual request histories,  completion
recovers clicked SIDs outside the query-only 7B top-100 list on 5.67\% 
of rows. A 14-day user-level online A/B test against 0.5B end-to-end
retrieval yields relative gains of 0.93\% in click-through rate and
1.05\% in revenue per thousand requests. 

\begin{samepage}
In summary, our key contributions are threefold: 

\noindent First, we identify shared SID-prefix structure and
query-dependent reuse boundaries in click logs, providing an empirical
basis for sharing large-model candidate planning across users while
retaining history-conditioned completion. 

\noindent Second, we develop ReuseGR to connect 7B offline planning with
0.5B online personalized retrieval through adaptive prefix-depth
selection, suffix-score distillation, and set-dependent branch admission. 

\noindent Third, we validate query--depth matching and personalized
completion through controlled experiments and comparisons with
complete-result caching and reranking, and demonstrate the system's
retrieval quality, serving efficiency, and business value on industrial
data and in a 14-day online A/B test. 
\end{samepage}

%% file: sections/02_related.tex
\section{Related Work}
\label{sec:related}

\paragraph{Generative retrieval and Semantic IDs.}
Generative retrieval learns candidate selection through identifier
prediction: entity names~\cite{decao2021genre}, document
IDs~\cite{tay2022transformer,wang2022nci}, or
substrings~\cite{bevilacqua2022seal}. Learned
identifiers~\cite{sun2023genret} and vector-quantized
codes~\cite{oord2017vqvae,lee2022rqvae} structure candidate representations. 
P5 unifies recommendation tasks as generation~\cite{geng2022p5}; TIGER
and LETTER construct semantic item codes~\cite{rajput2023recommender,wang2024letter}. 
OneRec and OneSearch extend this paradigm to unified recommendation and
search~\cite{onerec2024,chen2025onesearch,chen2026onesearchv2}. 
DGI jointly optimizes indexing and retrieval through a differentiable
training path and scaled cosine scores~\cite{wang2026dgi}. 

\paragraph{Search caching and offline generation.}
Precomputation reuses work across searches. Dense retrieval stores document
vectors~\cite{karpukhin2020dpr}; classical search precomputes score
bounds~\cite{ding2011blockmax} and caches results, postings, scores, and
list intersections~\cite{marktbase,ozcan2012fivelevel}, with index-aware
invalidation~\cite{list-cache}. For generative systems, RARE caches
commercial intentions in sponsored search~\cite{rare2024}; OneRec-Think
precomputes user-conditioned SID prefixes for online completion in
recommendation~\cite{liu2026onerecthink}. ReuseGR shares query-only SID
prefix sets across users and selects their depths from historical target
coverage, leaving final completion to each user's current history. 

\paragraph{Distillation and efficient online inference.}
Distillation transfers teacher distributions or
sequences~\cite{hinton2015distilling,kim2016seqkd}; SmartGR distills
hierarchical predictions and beam rankings~\cite{smartgr2024}, while
SID-MLP uses MLP prediction heads~\cite{sidmlp2024}. Parallel
prediction~\cite{hou2025rpg}, speculative
verification~\cite{speculative-decoding,lin2025atspeed}, and vectorized
constraints~\cite{su2026static} reduce decoding work. 
IO-aware attention~\cite{dao2022flashattention}, attention-state caching,
and memory management~\cite{kwon2023vllm,zheng2024sglang,zhao2026rcllm} improve
execution efficiency. ReuseGR combines large-model prefix reuse with distillation for the
smaller online suffix scorer. 

\paragraph{Budgeted decoding and policy optimization.}
Budgeted decoding allocates computation among candidate trajectories. 
GRC learns SID refinement trajectories and schedules reflection during
beam search~\cite{xing2026grc}. 
On-policy distillation supervises student-generated
sequences~\cite{agarwal2024gkd}; GRPO uses group-relative
rewards~\cite{grpo2024}. BudgetCut trains branch admission with these
objectives while keeping the planner and scorer fixed. 

%% file: sections/03_preliminary.tex
\section{Preliminaries}
\label{sec:preliminary}

\paragraph{Items, identifiers, and requests.}
\label{sec:pre-gr}
Let $\mathcal I$ be the item catalog. A fixed codebook maps item $i$ to
its complete Semantic ID (SID),
$y(i)=(y_1(i),\ldots,y_L(i))\in\{0,\ldots,V-1\}^L$. 
A SID layer is one token position. Token $y_t$ indexes a codeword; 
a prefix $y_{1:d}$ specifies a subtree. 
The bucket $\mathcal I(y)=\{i:y(i)=y\}$ contains items sharing a complete SID. 
Our codebook uses $L=4$ and $V=32$; the first two tokens define a coarse cell. 
We use the supplied item-to-SID mapping throughout all retrieval comparisons. 
Given request $x=(q,h)$, $q$ is the normalized query and $h$ is the
recent interaction history used by the retriever. 
A generator predicts full SIDs as
\begin{equation}
\pi_\theta(y\mid x)=\prod_{t=1}^{L}\pi_\theta(y_t\mid y_{<t},x).
\label{eq:autoregressive}
\end{equation}
Retrieval ranks distinct full SIDs; their buckets are expanded into items
for downstream ranking. Offline planning uses $q$ alone; online completion
uses the full $x$. Query-macro evaluation uses each query's history from its
first log row (canonical history); request-level evaluation uses each request's
actual history. 

\paragraph{Layer-wise reusability and KeyProbe.}
\label{sec:pre-reuse}
\label{sec:reusecut-r}
\label{sec:reusecut-d}
To assess which SID layers support query-level reuse, KeyProbe compares
query-specific and global token predictions. 
For target row $r$ with query $q$ and SID $y_r$, it removes the entire row
and predicts $y_{rt}$ by the most frequent token among other rows sharing
$(q,y_{r,<t})$. Denote this prediction by $\widehat y_{rt}^{q}$ and the
most frequent token across queries with the same prefix (the global mode)
by $\widehat y_{rt}^{0}$; 
ties select the smallest token.  Let $J_t(q)$ contain rows with a
leave-one-out query reference and
$\mathcal Q_t^+=\{q:|J_t(q)|>0\}$. Both predictors are evaluated on $J_t(q)$. 
For $z\in\{q,0\}$, define their per-query accuracy and its query-macro gain: 
\begin{equation}
\begin{aligned}
\widehat a_t^z(q)
&=\frac{1}{|J_t(q)|}\sum_{r\in J_t(q)}
\boldsymbol{1}[\widehat y_{rt}^{z}=y_{rt}],\\
\Delta R_t
&=\frac{1}{|\mathcal Q_t^+|}\sum_{q\in\mathcal Q_t^+}
\big[\widehat a_t^q(q)-\widehat a_t^0(q)\big]. 
\end{aligned}
\label{eq:rt-identifiable}
\end{equation}
Support is the fraction of all evaluation rows with a query reference; 
$\Delta R_t$ gives each query in $\mathcal Q_t^+$ equal weight. 
This diagnostic motivates query-level prefix sharing; historical prefix
coverage supplies ReuseCut's depth labels (Section~\ref{sec:reusecut}).
Separately, $D_t$ measures the 7B--0.5B next-token accuracy gap, given
the query and true preceding SID tokens, without history
(Appendix~\ref{app:diagnostics}). 

\paragraph{Reuse depth and serving cost.}
\label{sec:pre-split}
\label{sec:pre-latency}
\label{sec:pre-problem}
Depth $d\in\{0,\ldots,L-1\}$ is the number of cached SID tokens. A prepared
query leaves $L-d$ layers online; $d=0$ starts from the empty root. Mean online
latency comprises lookup time $t_0$, suffix evaluation, and branch selection
plus ranking the merged candidates with cost $t_r$: 
\begin{equation}
T(d)=t_0+\sum_{t=d+1}^{L}T_t(M_t;\theta)+t_r, 
\label{eq:latency-split}
\end{equation}
where $M_t$ counts branches scored at layer $t$. ReuseCut changes $d$. 
BudgetCut acts after the final-layer readout: it controls which scored
branches enter the final merged ranking and reduces its candidate pool. 

%% file: sections/04_method.tex
\section{Methodology}
\label{sec:method}

\subsection{System Overview}
\label{sec:reusegr}

\begin{figure*}[t]
\centering
\includegraphics[width=\textwidth]{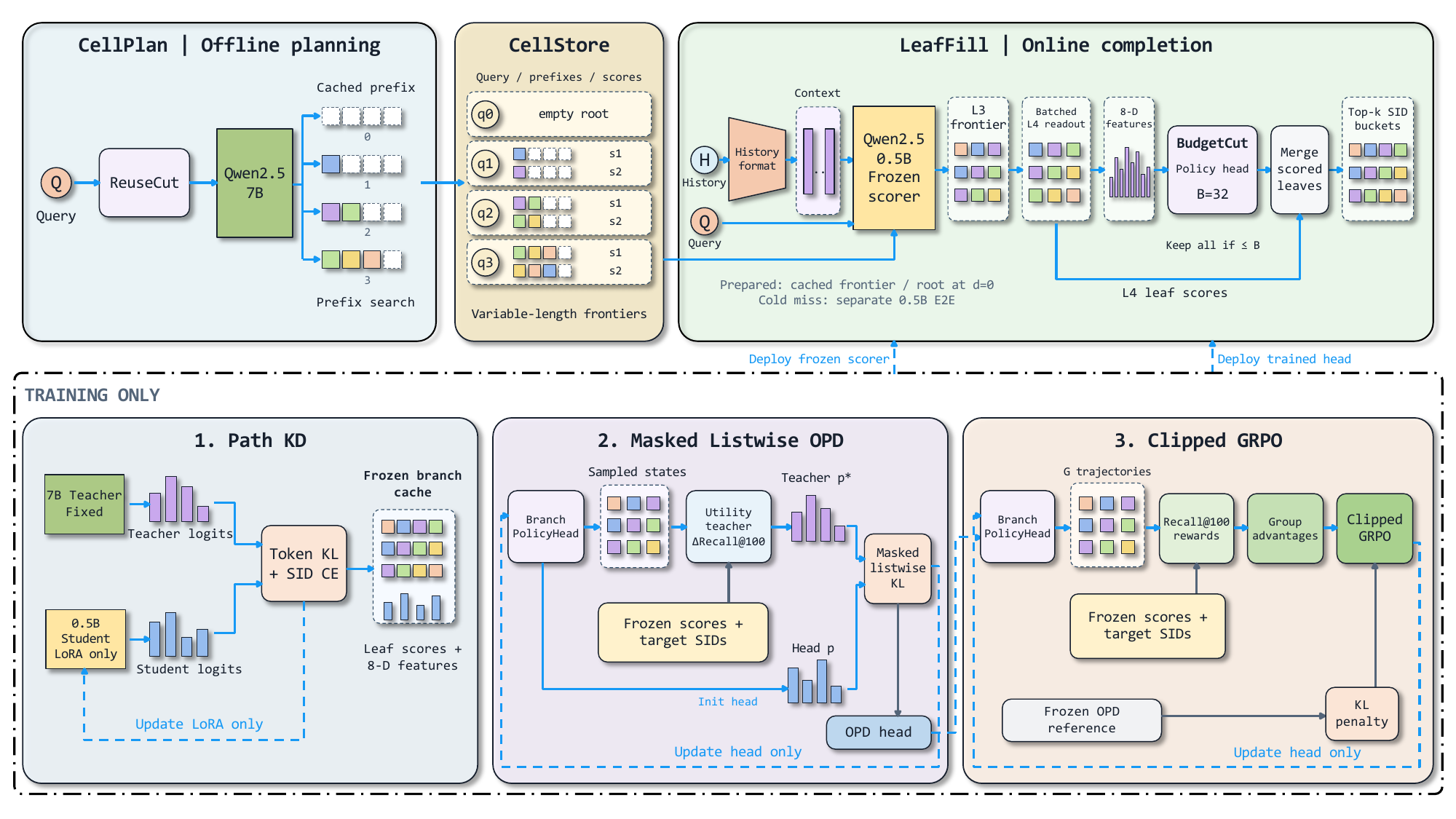}
\caption{ReuseGR architecture. Top: ReuseCut predicts cache depths, 
CellPlan plans prefixes offline, and CellStore shares them
with context-conditioned LeafFill. After batched $L_4$ readout, 
BudgetCut admits branches and merges their already-scored leaves
into top-$k$ SID buckets. Bottom: Path KD distills the 0.5B scorer before
freezing it; Masked Listwise OPD followed by Clipped GRPO trains only the
branch policy head. Cold misses use the separate 0.5B E2E endpoint. }
\Description{Top: offline planning shows cached prefixes of lengths zero, 
one, two, and three. A shared query-keyed cache feeds online completion, 
which formats the request history, expands to the layer-three frontier, 
scores layer-four leaves, extracts eight-dimensional branch features, 
admits at most B branches, and merges scored leaves into SID buckets. 
Prepared depth-zero queries start from the empty root; cold misses use a
separate 0.5B E2E model. Bottom: Path KD updates student LoRA; the frozen
scorer supports policy-head training with OPD and GRPO. }
\label{fig:overview}
\end{figure*}

ReuseGR separates shared candidate planning from personalized completion
(Figure~\ref{fig:overview}). Offline, a 7B model generates SID prefixes at
a depth selected for each query. Online, a 0.5B model expands these
prefixes using the current query and history; BudgetCut selects scored
branches for the final ranking. Cold misses use a separate 0.5B end-to-end (E2E) model.

\subsection{Adaptive Prefix Caching}
\label{sec:reusecut}
\label{sec:adaptive}

ReuseCut predicts cache depths from query text to balance reuse and
candidate coverage. Historical click-prefix sets supply depth labels; 
7B CellPlan independently generates the prefixes stored for completion. 

\paragraph{Model-generated prefixes.}
\label{sec:cellplan}
For query $q$ and depth $s\in\{1,\ldots,d_{\max}\}$, 7B CellPlan with
parameters $\theta^*$ selects the top-$N$ prefixes from $\mathcal V_s$, 
the depth-$s$ prefixes occupied by catalog items: 
\begin{equation}
\mathcal C_s(q)=\operatorname*{TopN}_{c\in\mathcal V_s}
\log\pi_{\theta^*}(c\mid q),
\label{eq:cellplan}
\end{equation}
Scores are cumulative masked log probabilities. With $N=32$, depth 1
retains all legal root tokens, depth 2 enumerates their legal children, 
and depth 3 uses exact best-first search. Each depth is ranked
independently for its top-$N$ frontier. 

\paragraph{Historical coverage labels. }
To label depths with sufficient historical target coverage, let $J(q)$
be the $n_q=|J(q)|$ aggregated outcome rows, with target SIDs $y_r$, 
for query $q$. Let
$\mathcal E_s^{(r)}(q)$ be the empirical reference set of
top-$N$ prefixes at depth $s$ ranked by historical click mass (computed via
whole-row leave-one-out on Fit and preceding-day logs on cross-day Test). The
empirical row-level coverage at depth $s$ is
\begin{equation}
\widehat C_s(q)=\frac{1}{n_q}\sum_{r\in J(q)}
\boldsymbol{1}[y_{r,1:s}\in\mathcal E_s^{(r)}(q)]. 
\label{eq:prefix-coverage}
\end{equation}
ReuseCut applies the Wilson-score lower endpoint ($z=1.96$)~\cite{wilson1927inference}
to $\widehat C_s(q)$ as a finite-support coverage criterion. For $n_q\ge5$, 
it assigns the supervision target $d^*(q)$ as the deepest level whose
coverage scores at every level through that depth meet threshold $\tau_C$: 
\begin{equation}
\begin{aligned}
d^*(q)=\max\big\{s\in\{0,\ldots,d_{\max}\}:\quad\\
\underline C_t(q)\geq\tau_C\ \text{for every }1\leq t\leq s\big\}, 
\end{aligned}
\label{eq:sstar}
\end{equation}
Actual reachability is measured on the model-generated sets
$\mathcal C_s(q)$. We set $d_{\max}=3$, $N=32$, and
$\tau_C=0.4$. Queries with $n_q < 5$ are
marked as \emph{Unknown}, assigned conservative depth $0$ during training-cache
construction, and excluded from depth-classifier fitting. 

\paragraph{Query-text depth prediction.}
ReuseCut predicts reusable depths from normalized query text, allowing
cache preparation for queries with insufficient click history. A
class-balanced multinomial logistic classifier is fitted on labeled Fit
queries. Let $\psi(q)$ encode hashed character 3--5-grams and length
features. For depth classes $\mathcal D=\{0,1,2,3\}$, it predicts
\begin{equation}
\widehat d(q)=\operatorname*{arg\,max}_{s\in\mathcal D}
\big(w_s^\top\psi(q)+b_s\big).
\label{eq:depth-predictor}
\end{equation}
Dev macro-F1 selects regularization, breaking ties by lower depth MAE
and then stronger regularization. Query-only prediction assigns
depths to unseen queries. 

\paragraph{Cache contents and availability. }
\label{sec:cellstore}
For an offline query pool $\mathcal Q$, CellStore
indexes each positive-depth entry as
\begin{equation}
q\longmapsto\big(\widehat d(q),\mathcal C_{\widehat d(q)}(q), 
\{s_q(c)\}_{c\in\mathcal C_{\widehat d(q)}(q)}\big), 
\label{eq:cellstore}
\end{equation}
where $s_q(c)=\log\pi_{\theta^*}(c\mid q)$ is the cached prefix score. 
Define $\mathcal C_0(q)=\{\varnothing\}$ and $s_q(\varnothing)=0$. 
Prepared queries with $\widehat d(q)=0$ use the Path KD scorer from this
root, without stored prefixes. Cold misses instead use the separate E2E
model. Fixed-SFT sets $d=2$ uniformly across cached queries. 

\subsection{Prefix-Conditioned Completion}
\label{sec:leaffill}

At cached depth $d$, 0.5B LeafFill with parameters $\theta$ completes
prefix $c\in\mathcal C_d(q)$ under request context $x=(q,h)$:
\begin{equation}
\pi_\theta(y_{d+1:L}\mid y_{1:d}=c,x). 
\label{eq:leaffill}
\end{equation}
All cache depths feed a common $L_3$ frontier of three-token branches
for $L_4$ scoring and BudgetCut selection. LeafFill encodes $(q,h)$ once, 
shares prompt KV states across branches, and restricts continuations to
the occupied codebook trie. For $d<3$, a best-first priority queue uses
the composite score below to form candidate set $\mathcal B_x$ at this
frontier; for $d=3$, cached $\mathcal C_3(q)$ supplies it directly. 
Score ties are broken lexicographically. Serving retains up to 64
branches: the top 56 by cumulative score plus up to eight deterministic
hash-selected alternatives. 

\paragraph{SID scoring.}
For a prefix $y_{1:t}$ extending cached $c=y_{1:d}$, define its
composite score by combining the offline prior and online suffix scores: 
\begin{equation}
s_x(y_{1:t})=\beta s_q(c)+
\sum_{j=d+1}^{t}\log\pi_\theta(y_j\mid y_{<j},x), 
\label{eq:scoring}
\end{equation}
for $d\le t\le L$. At $t=3$ this scores a branch; at $t=L$, $s_x(y)$
scores a complete SID. The weight $\beta$ balances offline and online
contributions: the adaptive decoder and matched depth controls use 1.0; 
Fixed-SFT uses its tuned value 1.5. Distinct SIDs are ranked by $s_x(y)$
to return top-$k$ candidate buckets. 

\subsection{Scorer Distillation}
\label{sec:distill}

Path KD strengthens online suffix scoring by matching token distributions
along true target prefixes. The frozen 7B teacher ($\mathrm T$) reads
$q_r$ alone; the 0.5B LoRA student ($\eta$) reads $(q_r,h_r)$. 
For Fit row $r$, split depth $d_r$ uses $d^*(q_r)$ with probability $0.5$, 
otherwise a uniform draw from $\{1,2\}$. 
For temperature $\tau$ and legal tokens $\mathcal V_{rt}$, define
$p^\eta_{rt}=\sigma_{\mathcal V_{rt}}(z^\eta_{rt}/\tau)$ and
$p^{\mathrm T}_{rt}=\sigma_{\mathcal V_{rt}}(z^{\mathrm T}_{rt}/\tau)$. 
Here $z$ denotes logits and $\sigma$ masked softmax. The distillation loss is
\begin{equation}
\begin{aligned}
\mathcal L_1(\eta)=\frac{1}{\sum_r a_r}
\sum_r\frac{a_r}{L-d_r}\sum_{t=d_r+1}^{L}
\tau^2\mathrm{KL}(p^{\mathrm T}_{rt}\Vert p^\eta_{rt}), 
\end{aligned}
\label{eq:leaf-distill}
\end{equation}
where $a_r$ is the query-normalized training-row weight and $\mathcal L_0(\eta)$
is supervised SID cross-entropy over active layers with observed or
click-derived soft targets. The combined objective is
\begin{equation}
\mathcal L(\eta)=(1-\alpha)\mathcal L_0(\eta)
+\alpha\mathcal L_1(\eta). 
\label{eq:student-loss}
\end{equation}
Only student LoRA~\cite{hu2022lora} parameters $\eta$ are optimized, at every
suffix layer. The teacher, base weights, and prefix entries stay fixed. 
We use $\tau=1$ and $\alpha=0.5$.
The distilled scorer is frozen before BudgetCut training. 

\subsection{BudgetCut: Recall-Aware Admission}
\label{sec:budgetcut}

BudgetCut optimizes recall of the merged top-100 ranking under a fixed
branch budget. High-scoring branches can introduce non-target leaves
that displace relevant SIDs. OPD and GRPO learn set-dependent admission: 
a branch's utility depends on those already selected. 
Freezing the planner and scorer preserves cached frontiers and leaf
scores, letting policy rollouts evaluate branch combinations without
repeated SID generation. 

The frozen 0.5B scorer evaluates valid $L_4$ continuations for all
three-token branches $a=y_{1:3}\in\mathcal B_x$ in one batch. 
At $d=3$, its at most $N=32\le B$ cached branches pass unchanged. Eight-dimensional features $g_x(a)$
combine branch scores, tree statistics, and top-8 leaf summaries
(Appendix~\ref{app:training}). All valid leaves ($\le32$ per branch) remain
available for final ranking. 

Conditioned on selected set $S$, the policy chooses from
$\mathcal A(S)=\mathcal B_x\setminus S$ until admitting
$B_x=\min(B,|\mathcal B_x|)$ branches without replacement. 
At $B=32$, this reduces the maximum merge pool from 2,048 to 1,024
already-scored leaves. 

Training caches branch features and legal leaf scores in $\mathcal F_x$. 
For top-100 distinct SIDs $\mathcal R_{100}(S;\mathcal F_x)$ in the selected
union, the reward and marginal recall utility are: 
\begin{align}
r_x(S) &= \frac{|\mathcal R_{100}(S;\mathcal F_x)\cap\mathcal Y(x)|}{|\mathcal Y(x)|}, \label{eq:reward} \\
U_x(b\mid S) &= r_x(S\cup\{b\})-r_x(S),\quad b\in\mathcal A(S), \label{eq:utility}
\end{align}
where $\mathcal Y(x)$ pools distinct full SIDs clicked in Fit for $(q,h)$, 
including targets outside the frontier. Policy training uses nonempty
groups. Utility is computed
after union re-ranking as a difference of Recall fractions, giving
$U_x\in[-1,1]$; negative values reflect target displacement. 
OPD initializes admission from these utilities; GRPO optimizes the complete set. 

A two-layer MLP with trainable parameters $\omega$ scores each available branch: 
\begin{equation}
f_\omega(a\mid S,x) = W_2\,\mathrm{GELU}(W_1\phi_x(a,S)+b_1)+b_2, 
\label{eq:policy-head}
\end{equation}
where $\phi_x(a,S)$ concatenates the eight-dimensional branch feature vector $g_x(a)$, the mean
feature vector of already-selected branches (zero when $S=\varnothing$), $|S|/B$, and $(|\mathcal B_x|-|S|)/|\mathcal B_x|$. 
The rollout action probability is: 
\begin{equation}
p_\omega(a\mid S,x)=\frac{\exp f_\omega(a\mid S,x)}
 {\sum_{a'\in\mathcal A(S)}\exp f_\omega(a'\mid S,x)}. 
\label{eq:budgetcut-policy}
\end{equation}
\subsubsection{Masked Listwise On-Policy Distillation}
\label{sec:opd}

At each visited selection state, OPD uses unscaled marginal utilities
$U_x(a\mid S)$ as teacher logits at unit temperature over $\mathcal A(S)$: 
\begin{equation}
p^*(a\mid S,x) = \frac{\exp U_x(a\mid S)}{\sum_{a'\in\mathcal A(S)}\exp U_x(a'\mid S)}. 
\label{eq:opd-teacher}
\end{equation}
The utility teacher supervises the branch head with the same action mask
$\mathcal A(S)$, using teacher-to-head KL divergence: 
\begin{equation}
\mathcal L_2(\omega) = \mathbb E_{x,S\sim d_{p_{\omega^-}}}\Big[
\sum_{a\in\mathcal A(S)}\bar p^*(a\mid S,x)\log\frac{\bar p^*(a\mid S,x)}{p_\omega(a\mid S,x)}\Big], 
\label{eq:opd}
\end{equation}
where $\operatorname{sg}$ stops gradients, $\bar p^*=\operatorname{sg}[p^*]$, 
and $d_{p_{\omega^-}}$ is the visited-state distribution of rollout
snapshot $\omega^-$ for each request $x$. 
Fit-only target-branch augmentation is specified in
Appendix~\ref{app:training}; Dev/Test candidates use no target labels. 

\subsubsection{Clipped Group-Relative Policy Optimization}
\label{sec:grpo}

GRPO~\cite{grpo2024} optimizes the quality of the complete admitted set $S$. 
For each request, $G$ trajectories are sampled from rollout policy $p_{\omega^-}$ with rewards
$r^{(g)}=r_x(S^{(g)})$ standardized into group advantages: 
\begin{equation}
A^{(g)}=\frac{r^{(g)}-\operatorname{mean}_{g'}r^{(g')}}{\operatorname{std}_{g'}r^{(g')}+\epsilon}.
\label{eq:advantage}
\end{equation}
Let $\rho_{gt}=\exp(\ell_{gt}^{\omega}-\ell_{gt}^-)$ be the importance ratio and $\kappa_{gt}$ the
divergence penalty relative to the frozen OPD reference policy $\bar\ell_{gt}$ (definitions in
Appendix~\ref{app:training}). The policy optimization objective is: 
\begin{equation}
\begin{aligned}
\mathcal L_3(\omega)=\mathbb E_{g,t}\Big[&-\min\big(\rho_{gt}\operatorname{sg}[A^{(g)}],\\
&\operatorname{clip}(\rho_{gt},1-\varepsilon,1+\varepsilon)
\operatorname{sg}[A^{(g)}]\big)\\
&+\lambda\kappa_{gt}\Big].
\end{aligned}
\label{eq:grpo}
\end{equation}
$\ell_{gt}^{\omega}$ and $\ell_{gt}^-$ denote chosen-action log probabilities under
the current and rollout heads, respectively; $g$ indexes trajectories
and $t$ selection steps. 
$\epsilon>0$ stabilizes group standardization; $\varepsilon$ is the
PPO~\cite{schulman2017ppo} clip parameter and $\lambda$ the reference-penalty
weight. OPD initializes the head for GRPO with the planner and scorer
frozen. Training samples actions; serving chooses the highest-probability
action at each step. 
\label{sec:training-order}
\label{sec:budgetcut-alg}

%% file: sections/05_experiments.tex
\section{Experiments}
\label{sec:experiments}

% Queue the main comparison and ablations near their experimental setup.
\begin{table*}[!t]
\centering
\caption{Query-macro complete-SID retrieval on 44,225 Test queries (canonical histories,
three-seed means). One A100, batch size 1; ReuseGR uses available prefixes.
Bold: column best. }
\label{tab:main}
\small
\setlength{\tabcolsep}{4.8pt}
\renewcommand{\arraystretch}{1.0}
\begin{tabular}{@{}lrrrrrrrr@{}}
\toprule
& \multicolumn{3}{c}{HR (\%)} & \multicolumn{3}{c}{Recall (\%)} & Latency & Speedup \\
\cmidrule(lr){2-4}\cmidrule(lr){5-7}
Method & @1 & @10 & @100 & @1 & @10 & @100 & (ms) $\downarrow$ & vs.\ 7B $\uparrow$ \\
\midrule
DSSM~\cite{huang2013dssm} & \TblVal{q.dssm.hr1} & \TblVal{q.dssm.hr10} & \TblVal{q.dssm.hr100} & \TblVal{q.dssm.rec1} & \TblVal{q.dssm.rec10} & \TblVal{q.dssm.rec100} & \TblVal{q.dssm.mean} & $\TblVal{q.dssm.speedup}\times$ \\
T5-DSI-SID~\cite{tay2022transformer} & \TblVal{q.t5.hr1} & \TblVal{q.t5.hr10} & \TblVal{q.t5.hr100} & \TblVal{q.t5.rec1} & \TblVal{q.t5.rec10} & \TblVal{q.t5.rec100} & \TblVal{q.t5.mean} & $\TblVal{q.t5.speedup}\times$ \\
0.5B E2E~\cite{qwen2025qwen25} & \TblVal{q.small.hr1} & \TblVal{q.small.hr10} & \TblVal{q.small.hr100} & \TblVal{q.small.rec1} & \TblVal{q.small.rec10} & \TblVal{q.small.rec100} & \TblVal{q.small.mean} & $\TblVal{q.small.speedup}\times$ \\
7B E2E (Teacher)~\cite{qwen2025qwen25} & \textbf{\TblVal{q.teacher.hr1}} & \textbf{\TblVal{q.teacher.hr10}} & \TblVal{q.teacher.hr100} & \TblVal{q.teacher.rec1} & \textbf{\TblVal{q.teacher.rec10}} & \textbf{\TblVal{q.teacher.rec100}} & \TblVal{q.teacher.mean} & $\TblVal{q.teacher.speedup}\times$ \\
SmartGR~\cite{smartgr2024} & \TblVal{q.smartgr.hr1} & \TblVal{q.smartgr.hr10} & \TblVal{q.smartgr.hr100} & \TblVal{q.smartgr.rec1} & \TblVal{q.smartgr.rec10} & \TblVal{q.smartgr.rec100} & \TblVal{q.smartgr.mean} & $\TblVal{q.smartgr.speedup}\times$ \\
OneRec-Think~\cite{liu2026onerecthink} & \TblVal{q.onerec-think.hr1} & \TblVal{q.onerec-think.hr10} & \TblVal{q.onerec-think.hr100} & \TblVal{q.onerec-think.rec1} & \TblVal{q.onerec-think.rec10} & \TblVal{q.onerec-think.rec100} & \TblVal{q.onerec-think.mean} & $\TblVal{q.onerec-think.speedup}\times$ \\
\midrule
\textbf{ReuseGR (Full)} & $\TblVal{q.full.hr1}$ & $\TblVal{q.full.hr10}$ & $\mathbf{\TblVal{q.full.hr100}}$ & $\mathbf{\TblVal{q.full.rec1}}$ & $\TblVal{q.full.rec10}$ & $\TblVal{q.full.rec100}$ & \textbf{\TblVal{q.full.mean}} & $\mathbf{\TblVal{q.full.speedup}\times}$ \\
\bottomrule
\end{tabular}
\end{table*}

\begin{table*}[!t]
\centering
\caption{Controlled ablations on Test. Upper: 7B planning, $\beta=1.0$ unless noted.
Lower: popularity (SFT, $\beta=1.0$) and 0.5B planning (fixed: SFT,
$\beta=1.5$; Full: Path KD and OPD+GRPO). Unpruned: up to 64 branches;
budgeted policies: $B=32$.}
\label{tab:ablation}
\label{tab:planner-ablation}
\small
\setlength{\tabcolsep}{4.8pt}
\renewcommand{\arraystretch}{1.0}
\begin{tabular}{@{}lrrrrrrr@{}}
\toprule
& \multicolumn{3}{c}{HR (\%)} & \multicolumn{3}{c}{Recall (\%)} & Latency \\
\cmidrule(lr){2-4}\cmidrule(lr){5-7}
Configuration & @1 & @10 & @100 & @1 & @10 & @100 & (ms) $\downarrow$ \\
\midrule
Fixed-SFT ($d=2$, $\beta=1.0$) & \TblVal{q.fixed2-b1.hr1} & \TblVal{q.fixed2-b1.hr10} & \TblVal{q.fixed2-b1.hr100} & \TblVal{q.fixed2-b1.rec1} & \TblVal{q.fixed2-b1.rec10} & \TblVal{q.fixed2-b1.rec100} & \textbf{\TblVal{q.fixed2-b1.mean}} \\
Fixed-SFT ($d=2$, $\beta=1.5$) & \TblVal{q.fixed2-b15.hr1} & \TblVal{q.fixed2-b15.hr10} & \TblVal{q.fixed2-b15.hr100} & \TblVal{q.fixed2-b15.rec1} & \TblVal{q.fixed2-b15.rec10} & \TblVal{q.fixed2-b15.rec100} & \textbf{\TblVal{q.fixed2-b15.mean}} \\
ReuseCut + SFT, unpruned & \TblVal{q.adaptive-sft.hr1} & \TblVal{q.adaptive-sft.hr10} & \TblVal{q.adaptive-sft.hr100} & \TblVal{q.adaptive-sft.rec1} & \TblVal{q.adaptive-sft.rec10} & \TblVal{q.adaptive-sft.rec100} & \TblVal{q.adaptive-sft.mean} \\
ReuseCut + Path KD, unpruned & \TblVal{q.adaptive-kd.hr1} & \TblVal{q.adaptive-kd.hr10} & \TblVal{q.adaptive-kd.hr100} & \TblVal{q.adaptive-kd.rec1} & \TblVal{q.adaptive-kd.rec10} & \TblVal{q.adaptive-kd.rec100} & \TblVal{q.adaptive-kd.mean} \\
ReuseCut + Path KD, score heuristic & \TblVal{q.heuristic.hr1} & \TblVal{q.heuristic.hr10} & \TblVal{q.heuristic.hr100} & \TblVal{q.heuristic.rec1} & \TblVal{q.heuristic.rec10} & \TblVal{q.heuristic.rec100} & \TblVal{q.heuristic.mean} \\
ReuseCut + Path KD + OPD & \TblVal{q.opd.hr1} & \TblVal{q.opd.hr10} & \TblVal{q.opd.hr100} & \TblVal{q.opd.rec1} & \TblVal{q.opd.rec10} & \TblVal{q.opd.rec100} & \TblVal{q.opd.mean} \\
\textbf{Full (OPD + GRPO)} & \textbf{\TblVal{q.full.hr1}} & \textbf{\TblVal{q.full.hr10}} & \textbf{\TblVal{q.full.hr100}} & \textbf{\TblVal{q.full.rec1}} & \textbf{\TblVal{q.full.rec10}} & \textbf{\TblVal{q.full.rec100}} & \TblVal{q.full.mean} \\
\midrule
Empirical popularity cache ($d=2$) & \TblVal{q.popularity.hr1} & \TblVal{q.popularity.hr10} & \TblVal{q.popularity.hr100} & \TblVal{q.popularity.rec1} & \TblVal{q.popularity.rec10} & \TblVal{q.popularity.rec100} & \TblVal{q.popularity.mean} \\
0.5B planner ($d=2$) & \TblVal{q.small-planner-fixed.hr1} & \TblVal{q.small-planner-fixed.hr10} & \TblVal{q.small-planner-fixed.hr100} & \TblVal{q.small-planner-fixed.rec1} & \TblVal{q.small-planner-fixed.rec10} & \TblVal{q.small-planner-fixed.rec100} & \TblVal{q.small-planner-fixed.mean} \\
0.5B planner + Full online & \TblVal{q.small-planner-full.hr1} & \TblVal{q.small-planner-full.hr10} & \TblVal{q.small-planner-full.hr100} & \TblVal{q.small-planner-full.rec1} & \TblVal{q.small-planner-full.rec10} & \TblVal{q.small-planner-full.rec100} & \TblVal{q.small-planner-full.mean} \\
\bottomrule
\end{tabular}
\end{table*}

% Queue the evidence figures and depth controls to keep all results in the main paper.

\begin{table*}[!t]
\centering
\caption{Depth-assignment controls on 44,225 Test queries with canonical
histories. Full: Path KD and BudgetCut ($B=32$); SFT: score-based decoding. 
Quality in \%; mean latency in ms on one A100, batch size 1. }
\label{tab:depth-matching}
\small
\setlength{\tabcolsep}{3.3pt}
\renewcommand{\arraystretch}{1.0}
\begin{tabular}{@{}llrrrrrrrr@{}}
\toprule
& & & & & \multicolumn{3}{c}{HR (\%)} & Recall & Mean \\
\cmidrule(lr){6-8}
System & Depth assignment & $\bar d$ & Steps & $C_{\mathrm h}$ & @1 & @10 & @100 & @100 & (ms) \\
\midrule
\textbf{Full} & \textbf{Adaptive} & \TblVal{adaptive.d} & \TblVal{adaptive.steps} & \TblVal{frontier-adaptive.ch} & \textbf{\TblVal{q.full.hr1}} & \textbf{\TblVal{q.full.hr10}} & \textbf{\TblVal{q.full.hr100}} & \textbf{\TblVal{q.full.rec100}} & \TblVal{q.full.mean} \\
Full & Fixed $d=2$ (retrained) & \TblVal{fixed2.d} & \TblVal{fixed2.steps} & \TblVal{frontier-fixed2.ch} & \TblVal{q.fixed2-full-retrained.hr1} & \TblVal{q.fixed2-full-retrained.hr10} & \TblVal{q.fixed2-full-retrained.hr100} & \TblVal{q.fixed2-full-retrained.rec100} & \TblVal{q.fixed2-full-retrained.mean} \\
Full & Fixed $d=2$ (inference switch) & \TblVal{fixed2.d} & \TblVal{fixed2.steps} & \TblVal{frontier-fixed2.ch} & \TblVal{q.fixed2-full-switch.hr1} & \TblVal{q.fixed2-full-switch.hr10} & \TblVal{q.fixed2-full-switch.hr100} & \TblVal{q.fixed2-full-switch.rec100} & \TblVal{q.fixed2-full-switch.mean} \\
Full & Permuted depths & \TblVal{adaptive.d} & \TblVal{adaptive.steps} & \TblVal{frontier-permuted.ch} & \TblVal{q.permuted-full.hr1} & \TblVal{q.permuted-full.hr10} & \TblVal{q.permuted-full.hr100} & \TblVal{q.permuted-full.rec100} & \TblVal{q.permuted-full.mean} \\
\midrule
SFT & Adaptive & \TblVal{adaptive.d} & \TblVal{adaptive.steps} & \TblVal{frontier-adaptive.ch} & \TblVal{q.adaptive-sft.hr1} & \TblVal{q.adaptive-sft.hr10} & \TblVal{q.adaptive-sft.hr100} & \TblVal{q.adaptive-sft.rec100} & \TblVal{q.adaptive-sft.mean} \\
SFT & Permuted depths & \TblVal{adaptive.d} & \TblVal{adaptive.steps} & \TblVal{frontier-permuted.ch} & \TblVal{q.permuted-sft.hr1} & \TblVal{q.permuted-sft.hr10} & \TblVal{q.permuted-sft.hr100} & \TblVal{q.permuted-sft.rec100} & \TblVal{q.permuted-sft.mean} \\
\bottomrule
\end{tabular}
\end{table*}

We evaluate overall performance (RQ1), adaptive prefix reuse (RQ2), scorer
and policy contributions (RQ3), personalization and generalization (RQ4), 
serving cost (RQ5), and online outcomes (RQ6). 

\subsection{Experimental Setup}
\label{sec:exp-setup}

\paragraph{Data and evaluation.}
AE-PV is an industrial search-advertising dataset with 3.38M catalog items, 
indexed by four-layer SIDs with $V=32$. The Day-1 modeling subset has
1,000,042 rows across 43,776 queries, split by query into Fit (41,651) 
and Dev (2,125). 
Sampling targets come from cross-day overlap in the full PV8M logs: 
20.97\% of queries, 36.87\% of logged rows, and 37.68\% of clicks. 
Relative to the modeling subset, the day-2 source has 9,276 Seen queries
among 377,532 (2.46\%). We retain these and select whole Unseen query
groups to match the target shares. Test contains 992,118 rows and 44,225 queries: 
9,276 Seen and 34,949 Unseen. Each query has 15.77 distinct clicked SIDs on average. 
The main evaluation uses the first source-row history $\bar h_q$ and prepared
prefixes for both Seen and Unseen queries. Training details appear in
Appendix~\ref{app:implementation}. 

\paragraph{Baselines. }
We compare three generative retrieval routes and an index-based reference. 
\emph{Full online generation: } T5-DSI-SID adapts DSI with a
T5-Base encoder--decoder and valid-trie constraints to generate four-token
SIDs~\cite{raffel2020t5,tay2022transformer}. The all-online 0.5B and 7B
end-to-end (E2E) baselines use Qwen2.5-0.5B-Instruct and Qwen2.5-7B-Instruct, 
respectively~\cite{qwen2025qwen25}; both generate all four SID tokens
online from the query and history. 

\emph{Distillation-based acceleration: } SmartGR transfers hierarchical
SID predictions and beam rankings from 7B to 0.5B while retaining
four-token online decoding~\cite{smartgr2024}. 

\emph{Prefix precomputation: } our search adaptation of
OneRec-Think~\cite{liu2026onerecthink} caches two-token prefixes per query
from $(q,\bar h_q)$ and uses 0.5B to complete them under $(q,h)$. 

DSSM~\cite{huang2013dssm} is the index-based reference: it retrieves 1,000
items by approximate nearest-neighbor search, then maps and deduplicates
them to top-100 SIDs. All methods share the catalog, item-to-SID mapping, 
and Test targets. ReuseGR combines 7B planning and 0.5B completion with
ReuseCut ($\tau_C=0.4$, $N=32$, $\beta=1.0$), Path KD ($\alpha=0.5$), 
and BudgetCut ($B=32$). Fixed-depth controls are evaluated in the ablations. 

\paragraph{Query-level metrics.}
Let $\mathcal Y_q$ pool distinct clicked SIDs over query $q$'s Test rows,
and $\mathcal R_k(q)$ be its top-$k$ distinct predictions. Queries receive equal weight: 
\begin{equation}
\begin{aligned}
\mathrm{HR@}k&=\frac{1}{|\mathcal Q|}\sum_q
\boldsymbol{1}[\mathcal R_k(q)\cap\mathcal Y_q\ne\varnothing],\\
\mathrm{Recall@}k&=\frac{1}{|\mathcal Q|}\sum_q
\frac{|\mathcal R_k(q)\cap\mathcal Y_q|}{|\mathcal Y_q|}. 
\end{aligned}
\label{eq:metrics}
\end{equation}
A query can have multiple ground truths. Hit@k is 1 if any is retrieved
(0 otherwise); Recall@k is the fraction of ground truths retrieved. 
We report percentages before item expansion at $k\in\{1,10,100\}$ and
percentage-point differences. Model results average seeds $\{0,17,42\}$
per query; paired-bootstrap 95\% intervals resample these query means
2,000 times~\cite{efron1979bootstrap}. 

\paragraph{Timing boundary.}
Timings use 44,225 canonical queries on one A100, batch size 1, with
shared prompt/prefix KV states and batched branch scoring. 
ReuseGR uses prepared prefixes; cold misses run 0.5B E2E. 
Mixed-traffic latency estimates are reported separately. 

\subsection{Overall Performance (RQ1)}
\label{sec:exp-main}
\label{sec:exp-tradeoff}

ReuseGR reaches 85.82\% HR@100 and 46.05\% Recall@100 at 71.45\,ms
(Table~\ref{tab:main}), gaining 3.83 HR@100 and 3.40 HR@10 points
($p<0.001$) over 0.5B E2E with 38.9\% lower mean latency. 
Against the distilled SmartGR baseline, it gains 2.26 HR@100
points and reduces mean latency by 37.4\%. 

Against OneRec-Think, ReuseGR gains 4.34 HR@100 points with
3.7\% lower mean latency. Its HR@100 is comparable to 7B E2E
(85.82/85.71\%) with a $2.36\times$ speedup; SID Recall@100 is
46.05/50.96\% and Item Recall@500 is 44.68/49.10\%, respectively. 

\paragraph{SID-to-item retrieval.}
\label{sec:exp-item-expansion}
An in-memory compressed sparse row (CSR) index expands top-100 SIDs
into disjoint item buckets under matched quotas of 100/500 items, adding
0.35\,ms. ReuseGR reaches 80.64/85.36\% Item HR@100/500 versus
76.42/81.54\% for 0.5B E2E; Item Recall@500 rises from 43.92\% to 44.68\%. 
Full item metrics and the Dev-selected expansion rule appear in
Appendix~\ref{app:additional-results}. 

\begin{figure}[t]
\centering
\includegraphics[width=\linewidth]{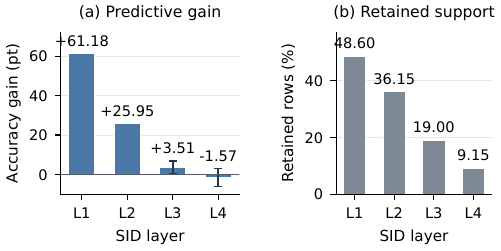}
\caption{Layer-wise KeyProbe: (a) query-macro accuracy gain over the
matched global-mode baseline; (b) retained row share. Error bars give
the reported 95\% intervals for layers 3/4. }
\Description{The first panel shows query-key predictive accuracy gains of 61.18, 25.95, 3.51, and minus 1.57 percentage points across four SID layers. The second panel shows retained row shares of 48.60, 36.15, 19.00, and 9.15 percent of evaluation rows. }
\label{fig:keyprobe}
\end{figure}

\subsection{Adaptive Prefix Reuse (RQ2)}
\label{sec:exp-mechanism}

\paragraph{Layer-wise evidence.}
\label{sec:exp-reusability}
\label{sec:exp-learnable}

On the rows retained at each layer, KeyProbe's query-key gain
(Eq.~\ref{eq:rt-identifiable}; Figure~\ref{fig:keyprobe}) is
61.18/25.95 points at layers 1/2 and 3.51 at layer 3; the layer-4
interval spans zero. This motivates shared coarse prefixes and
history-conditioned completion. Given the query and true preceding SID
tokens, 7B improves next-token accuracy over 0.5B by 1.69--2.53 points
without history, motivating suffix distillation
(Appendix~\ref{app:diagnostics}).

\paragraph{Depth and candidate reachability.}
\label{sec:exp-depth}
ReuseCut assigns 7,446, 11,374, 11,679, and 13,726 Test queries to depths
0--3, averaging 2.28 online steps. Let $C_{\mathrm h}$ be the fraction of
queries with any target SID reachable through the cached frontier, and
$C_{\mathrm r}$ the mean fraction of reachable target SIDs per query. 
These upper-bound HR and Recall before branch admission and the top-$k$
cutoff. Adaptive reuse raises $(C_{\mathrm h},C_{\mathrm r})$ from
$(96.69,74.18)\%$ at fixed $d=2$ to $(97.84,78.62)\%$. 

Table~\ref{tab:depth-matching} compares Full depth controls using 7B
planning, Path KD, and BudgetCut ($B=32$). The retrained fixed-$d=2$
control matches Full's training budget; the inference switch keeps Full's weights. 
Adaptive Full trades 4.20\,ms additional mean latency for gains of
1.67 HR@100 and 0.65 Recall@100 points over the retrained control. 

Depth permutation preserves depth counts and the 2.28 average online
steps, using query-specific frontiers and the Full scorer/head. Adaptive
assignment gains 2.08 $C_{\mathrm h}$ points and 3.17/2.20
HR@100/Recall@100 points, demonstrating the benefit of query--depth matching. 

Under SFT, adaptive assignment gains 1.20 HR@100 points over permuted
depths with a 0.25-point Recall@100 trade-off; Full improves both metrics. 
Appendix~\ref{app:coverage} gives
uniform SFT references and depth-label counts. On 6,237
support-qualified Test-Seen queries ($n_q\ge5$), depth prediction reaches
68.91\% exact accuracy, 95.13\% within one layer, and MAE 0.36. 

\paragraph{Source of reusable prefixes. }

With SFT completion at $d=2$ and $\beta=1.0$, 7B planning reaches
82.04\% HR@100 versus 76.48\% for empirical popularity with global Fit
fallback, a 5.56-point gain. At $\beta=1.5$, 7B/0.5B planning reaches
82.11/81.82\%; with identical Full online components, it reaches
85.82/83.46\% (Table~\ref{tab:planner-ablation}). These controls show gains
from semantic planning and the larger planner. 

\subsection{Component and Policy Ablations (RQ3)}
\label{sec:exp-ablation}
\label{sec:exp-extensions}

For suffix scoring, Path KD improves all six metrics on the same
adaptive, unpruned frontier at unchanged latency: HR@100 rises from
82.64\% to 83.32\% and Recall@100 from 44.38\% to 44.76\%
(Table~\ref{tab:ablation}). For branch selection, BudgetCut (OPD+GRPO) admits up to 32 branches
after $L_4$ scoring. With the scorer and $B=32$ fixed, it gains 2.91 HR@100
and 1.53 Recall@100 points over the score heuristic for 0.42\,ms
additional latency. Reachability remains
$(C_{\mathrm h},C_{\mathrm r})=(97.84,78.62)\%$
throughout. 

Against the unpruned union, BudgetCut merges fewer already-scored
leaves, saving 4.64\,ms and gaining 2.50 HR@100 points (95\% CI
$[2.22,2.78]$). The score heuristic saves 5.06\,ms with a 0.41-point
HR@100 trade-off. Pruning requires $|\mathcal B_x|>B$; depth-3 queries
pass unchanged. Full gains 3.71 HR@100 points over Fixed-SFT ($d=2$, $\beta=1.5$)
for 2.12\,ms additional mean latency. 

\begin{figure}[t]
\centering
\includegraphics[width=\linewidth]{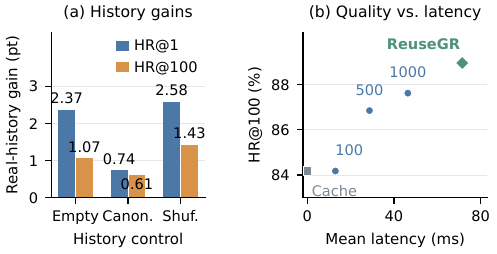}
\caption{Personalization on 286,046 non-canonical history rows. 
(a) Real-history gains over history controls. 
(b) Row-macro HR@100 versus canonical-workload mean latency (A100, batch size 1). 
Numeric labels: reranking pool size; Cache: 100 complete SIDs.
ReuseGR uses available prefixes. }
\Description{The left panel shows real-history HR at 1 and HR at 100 gains over empty, canonical, and within-query shuffled histories. The right panel compares a 7B complete-result cache, 0.5B reranking of 100, 500, or 1000 cached SIDs, and ReuseGR on HR at 100 and mean latency. }
\label{fig:personalization}
\end{figure}

\subsection{Personalization and Generalization (RQ4)}
\label{sec:exp-stratification}

We test history-dependent prefix completion and target recovery beyond
cached complete-SID lists. We evaluate 286,046 non-canonical history rows from 12,454 multi-history
queries, weighting rows equally against their clicked SIDs
(Table~\ref{tab:request-cache-history}; Figure~\ref{fig:personalization}). 
Query-only 7B caches top-$M$ complete SIDs, optionally reranked by 0.5B
with history. ReuseGR uses real, first-row canonical $\bar h_q$, empty, 
or within-query shuffled histories. 

\begin{table}[!b]
\centering
\caption{History-conditioned personalization. HR: three-seed row-macro
(\%, 286,046 rows); mean latency: 44,225 canonical queries, 
one A100, batch size 1. Cache sizes are payloads. }
\label{tab:request-cache-history}
\small
\setlength{\tabcolsep}{2.8pt}
\renewcommand{\arraystretch}{1.0}
\begin{tabular}{@{}lrrrr@{}}
\toprule
Method & HR@1 & HR@100 & ms & Bytes/query \\
\midrule
7B cache, $M=100$ & \TblVal{row.cache100.hr1} & \TblVal{row.cache100.hr100} & \TblVal{row.cache100.mean} & \TblVal{row.cache100.bytes} \\
Rerank, $M=100$ & \TblVal{row.rerank100.hr1} & \TblVal{row.rerank100.hr100} & \TblVal{row.rerank100.mean} & \TblVal{row.rerank100.bytes} \\
Rerank, $M=500$ & \TblVal{row.rerank500.hr1} & \TblVal{row.rerank500.hr100} & \TblVal{row.rerank500.mean} & \TblVal{row.rerank500.bytes} \\
Rerank, $M=1000$ & \TblVal{row.rerank1000.hr1} & \TblVal{row.rerank1000.hr100} & \TblVal{row.rerank1000.mean} & \TblVal{row.rerank1000.bytes} \\
\midrule
ReuseGR, empty & \TblVal{row.empty.hr1} & \TblVal{row.empty.hr100} & \TblVal{row.empty.mean} & \TblVal{row.empty.bytes} \\
ReuseGR, canonical & \TblVal{row.canonical.hr1} & \TblVal{row.canonical.hr100} & \TblVal{row.canonical.mean} & \TblVal{row.canonical.bytes} \\
ReuseGR, shuffled & \TblVal{row.shuffled.hr1} & \TblVal{row.shuffled.hr100} & \TblVal{row.shuffled.mean} & \TblVal{row.shuffled.bytes} \\
\textbf{ReuseGR, real} & \textbf{\TblVal{row.real.hr1}} & \textbf{\TblVal{row.real.hr100}} & \TblVal{row.real.mean} & \TblVal{row.real.bytes} \\
\bottomrule
\end{tabular}
\end{table}

With cache and weights fixed, real histories improve HR@100 by
1.07/0.61/1.43 points over empty/canonical/shuffled histories. 
HR@1 gains over canonical/shuffled histories are 0.74/2.58 points; 
29.4\% of top-1 SIDs change versus canonical input. 

At $M=100$, history-aware reranking retains the cached top-100 SID set. 
Expanding the pool to 1,000 raises HR@100 from 84.18\% to 87.62\%; 
ReuseGR reaches 88.95\% with a 256-byte prefix payload at 71.45\,ms. 

ReuseGR retrieves the clicked SID outside the 7B Top-100 list on 5.67\% 
of rows, recovering 35.84\% of rows missed by that list. Shared hits (83.28\%) and
cache-only hits (0.90\%) yield a net 4.77-point HR@100 gain. 

\begin{figure}[t]
\centering
\includegraphics[width=\linewidth]{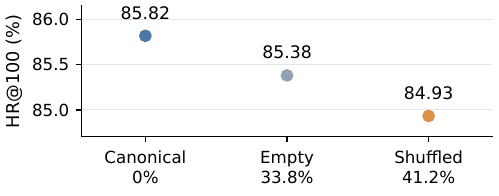}
\caption{History sensitivity on 44,225 canonical-history Test queries
(RQ4). Points show query-macro HR@100; percentages below the history
settings give top-1 SID change rates relative to canonical input. }
\Description{Canonical, empty, and shuffled histories yield HR at 100 of
85.82, 85.38, and 84.93 percent. Their top-one SID change rates relative
to canonical input are zero, 33.8, and 41.2 percent.}
\label{fig:robustness}
\label{tab:history-perturb}
\end{figure}

On the full 44,225-query canonical-history workload, histories improve
query-macro HR@100 by 0.44/0.89 points over empty/shuffled input
(Figure~\ref{fig:robustness}). For query generalization, ReuseGR with
prepared prefixes raises Unseen HR@10 from 60.46\% to 63.01\% over
0.5B E2E; Head/Torso mean latencies are 55.24/68.14\,ms
(Table~\ref{tab:stratification-full}). 

\subsection{Cache Availability and Cost (RQ5)}
\label{sec:exp-serving}
\label{sec:exp-cache-replay}
\label{sec:exp-amortization}
\label{sec:discussion}
\label{sec:deployment}
\label{sec:query-adaptive}
\label{sec:uncached}

\begin{table}[!t]
\centering
\caption{Serving scenarios. Q/P: prepared shares by queries/logged rows
(\%). Static Q uses measured partition quality; starred rows project
query-macro endpoints. All mixed mean latencies are projections. }
\label{tab:cache-replay}
\small
\setlength{\tabcolsep}{3.2pt}
\renewcommand{\arraystretch}{1.0}
\begin{tabular}{@{}lrrrr@{}}
\toprule
Regime & Q & P & HR@100 & Mean (ms) \\
\midrule
Prepared & \TblVal{serving.prepared.qshare} & \TblVal{serving.prepared.pshare} & \TblVal{q.full.hr100} & \TblVal{q.full.mean} \\
Cold (0.5B E2E) & \TblVal{serving.cold.qshare} & \TblVal{serving.cold.pshare} & \TblVal{q.small.hr100} & \TblVal{q.small.mean} \\
Static Q & \TblVal{serving.static-q.qshare} & \TblVal{serving.static-q.pshare} & \TblVal{serving.static-q.hr100} & \TblVal{serving.static-q.mean} \\
Static mix$^*$ & \TblVal{serving.static-mix.qshare} & \TblVal{serving.static-mix.pshare} & \TblVal{serving.static-mix.hr100} & \TblVal{serving.static-mix.mean} \\
Backfill$^*$ & \TblVal{serving.backfill.qshare} & \TblVal{serving.backfill.pshare} & \TblVal{serving.backfill.hr100} & \TblVal{serving.backfill.mean} \\
Rolling 1M$^*$ & \TblVal{serving.rolling.qshare} & \TblVal{serving.rolling.pshare} & \TblVal{serving.rolling.hr100} & \TblVal{serving.rolling.mean} \\
\bottomrule
\end{tabular}
\end{table}

For partial cache availability, Static Q serves 9,276 Seen queries with
ReuseGR and 34,949 Unseen queries
with 0.5B E2E, reaching 82.38\% query-macro HR@100 and
45.70\% Recall@100 (Table~\ref{tab:cache-replay}). Backfill/Rolling
project fixed query-macro quality and latency endpoints against prepared
logged-row share $p$, including cached prefixes and $d=0$ roots. 
Of 44,225 prepared queries, 36,779 (83.16\%) have positive depth. 
For maintenance, prefix payloads total 256\,MB per 1M entries; 10\% daily refresh costs 2.44 A100
GPU-hours. Two cache hits cover the planning cost. 

\begin{samepage}
\subsection{Online Evaluation (RQ6)}
\label{sec:exp-online}

We conducted a 14-day online A/B test on a leading global e-commerce platform serving hundreds of millions of users. 

ReuseGR replaces the 0.5B E2E retriever. ReuseCut
selects a cache depth per query, and the 7B planner stores SID prefixes
and scores. Prepared queries use the current history to complete cached
prefixes or the depth-zero root; BudgetCut filters branches before top-100 SID
ranking and item expansion. 
\par
\end{samepage}

Request serving and cache maintenance run separately. A cache miss
immediately uses the original 0.5B E2E model to decode all four SID layers
with the current context. Missed queries with positive predicted depth
enter a deduplicated, rate-limited queue for asynchronous 7B planning. 
Published entries serve later requests; periodic refresh limits staleness. 
Depth-zero misses remain fully online without backfill. Requests therefore
never wait for 7B planning, and both hit and miss paths retain current
user context. 

A 14-day user-level A/B test assigned 10\% traffic per arm in production
search advertising. Other retrieval channels, quotas, ranking, and
bidding were unchanged. From pooled counts, CTR (clicks/impressions) and
RPM (revenue per 1,000 requests) rose 0.93\% and 1.05\% relative to
baseline. Orders per request rose 0.78\%; CVR (orders/clicks) fell 0.21\%. 
SID-tagged impression/click shares rose 3.00/4.70 percentage points; the
share of requests with a SID-tagged impression rose 5.36 points. These
shares measure channel participation with potentially overlapping tags.

%% file: sections/06_discussion.tex
% Production-serving content is integrated into RQ5 in sections/05_experiments.tex.

%% file: sections/07_conclusion.tex
\section{Conclusion}
\label{sec:conclusion}

ReuseGR combines 7B offline query-prefix planning with 0.5B online
history-conditioned completion, using ReuseCut for adaptive depth, 
Path KD for suffix scoring, and BudgetCut for branch admission. Adaptive prefix reuse balances retrieval quality and efficiency. 
On AE-PV's 3.38M-item catalog, 
ReuseGR achieves 85.82\% HR@100 and 46.05\% Recall@100 at 71.45\,ms with
available prefix entries: 38.9\% lower latency than 0.5B e2e and comparable
HR@100 to 7B e2e. Query-level prefix reuse shares computation across
requests, while each request's history guides candidate selection within
the cached subtrees.
A 14-day online A/B test yields relative CTR and RPM gains of 0.93\% and
1.05\% over the 0.5B e2e baseline.

%% file: sections/08_appendix.tex
\appendix

\section{Experimental Protocol and Reproducibility}
\label{app:dataset}
\label{app:implementation}

\paragraph{Data and sampling.}
Normalization trims, lowercases, and collapses whitespace. 
Full PV8M logs, referenced to full day-1 logs, supply cross-day targets of
20.9746/36.8653/37.6813\% by queries/logged rows/click mass.
Sampled Seen queries overlap the 43,776-query Day-1 modeling subset. 
Sampling (seed 0) retains all 9,276 overlaps from 8,997,726 day-2 rows
(377,532 queries), then selects whole Unseen groups to match targets
within $10^{-6}$. Test has 992,118 clicked-SID outcome rows and 1,014,759
clicks: 975,968 rows have one click; 16,150 have multiple clicks. 
Each row has one scalar item ID and complete SID; its click count records
repeated clicks on that item.
Row-macro metrics weight rows equally; prefix frequencies use click mass; 
mixture scenarios use logged-row shares. Canonical contexts use each query's
first source-row history; $\mathcal Y_q$ deduplicates its Test target SIDs. 
Day-1 Fit (41,651 queries, 947,519 rows) trains both SID models; 
Dev (2,125 queries, 52,523 rows) selects them. Fit/Dev and Dev/Test are
query-disjoint; all 9,276 Seen Test queries occur in Fit. 

The catalog occupies 1,001 coarse cells and 355,438 leaf SIDs
(9.51 items per leaf). Test queries average 15.77 distinct clicked SIDs
and 19.50 items. 

\paragraph{Prompts and history.}
Offline CellPlan uses \texttt{Query: \{norm\_q\}\allowbreak\textbackslash nPredict target SID:},
with query input alone and no reasoning tokens. 
Online LeafFill uses \texttt{Query: \{norm\_q\}\allowbreak\textbackslash nHistory: \{h\_slots\}\allowbreak\textbackslash nPredict target SID:}
(192-token limit). History packs up to 20 recent interactions, most-recent-first,
as \texttt{"\{title\_trunc\} [\{s1\}.\{s2\}.\{s3\}.\{s4\}]"}. 
Real histories contain only interactions before the request timestamp. 
Prepared queries filter history by the offline 7B top-1 $L_1$ codeword. 
The 0.5B E2E baseline and cold fallback use the logged query-match mask
(\texttt{hist\_qmatch}), then four-layer trie-constrained decoding (beam 32). 
Shuffling samples a different history from the same query, with seed 42. 

\paragraph{Models and optimization.}
Qwen2.5-7B-Instruct and Qwen2.5-0.5B-Instruct~\cite{qwen2025qwen25} use four SID layers
with 32 values each. Path KD's supervised term uses targets $u_{rt}$: 
\begin{equation}
\mathcal L_0(\eta)=\frac{1}{\sum_r a_r}\sum_r\frac{a_r}{L-d_r}
\sum_{t=d_r+1}^{L}\mathrm{CE}(u_{rt},p_{rt}(\eta)), 
\end{equation}
For row click count $c_r$ and Fit rows $J(q_r)$ of query $q_r$, 
$a_r=c_r/\sum_{r'\in J(q_r)}c_{r'}$; thus each query's weights sum to one. 
Each epoch visits every Fit row once in a query-stratified shuffle, 
giving equal total objective weight per query. 
$p_{rt}(\eta)$ is the validity-masked, temperature-1 student distribution. 
For each $(q_r,y_{r,<t})$, soft targets normalize Fit click frequencies
over legal next tokens, with zero mass outside $\mathcal V_{rt}$. 
They are blended equally with the observed one-hot target before
combining supervised and distillation losses. 
KD uses AdamW, learning rate $2.5\times10^{-4}$, batch size 96, eight epochs, 
and LoRA rank/alpha/dropout 64/128/0.05 on
attention and feed-forward projections. Main and depth-control experiments
use model-based planning, with click-prior adjustments and history reranking
bonuses disabled. Full HR@100 has across-seed standard deviation 0.09
points. Anonymized records retain normalized queries, history
signatures, SIDs, and click counts without personally identifiable information; 
commercial logs remain private under corporate data governance. 

\paragraph{Empirical popularity control.}
At $d=2$, Fit click counts rank query-specific coarse cells. Global Fit
frequency fills unused slots up to 32 and supplies all 32 cells for queries
without Fit clicks. Prefix scores are add-one-smoothed log frequencies
($\beta=1.0$); ties follow coarse-cell ID. Completion uses the fixed-depth
0.5B SFT scorer and score-based decoder. The matched 7B control also uses
$d=2$ and $\beta=1.0$. 

\paragraph{Baseline configurations.}
DSSM averages the first 32 WordPiece embeddings, applies
$512\to256\to128$ MLP towers and $L_2$ normalization, and trains on Fit
click pairs for five epochs with InfoNCE temperature 0.05. Retrieval uses
cosine scores and Faiss HNSW; each SID takes its maximum item score. 
SmartGR uses Qwen2.5-7B/0.5B teacher/student and beam 32, training on Fit
for eight epochs with AdamW at $2\times10^{-4}$. Hierarchy and
beam-ranking distillation terms each have weight 0.5; Dev HR@10 selects
the checkpoint. The OneRec-Think search adaptation caches 7B top-32
$d=2$ prefixes by query, using $(q,\bar h_q)$ offline. Online 0.5B
completes two layers under $(q,h)$ with beam 32; timing includes lookup
and both steps. 

\section{Additional Prefix-Reuse Analyses}
\label{app:diagnostics}
\label{app:coverage}

\paragraph{Layer-wise diagnostics.}
For valid target rows $\mathcal J_t$, the next-token model-scale gain is
\begin{equation}
\begin{aligned}
D_t=\frac{1}{|\mathcal J_t|}\sum_{r\in\mathcal J_t}
\Big(&\boldsymbol{1}[\widehat y^{(1)}_{rt}=y_{rt}]\\
&-\boldsymbol{1}[\widehat y^{(0)}_{rt}=y_{rt}]\Big), 
\end{aligned}
\label{eq:dt}
\end{equation}
where 7B/0.5B predictions $\widehat y^{(1)}_{rt}/\widehat y^{(0)}_{rt}$ condition
on query $q_r$ and true $y_{r,<t}$, without history. All predictors use 992,118 equally
weighted rows per layer, separate from KeyProbe's subsets. Gains use
unrounded accuracies; 2,000 paired query-cluster resamples retain all rows. 
At layer 4, 7B exceeds the global-mode predictor by 11.50 points
(95\% CI $[10.41,12.56]$). Effective vocabulary and random accuracy are
aggregated separately. 

\begin{table}[!htbp]
\centering
\caption{KeyProbe support and query-key predictive gain by SID layer.
Support is retained row share (\%); gain and 95\% intervals are query-macro
accuracy differences (percentage points). }
\label{tab:reusability}
\small
\setlength{\tabcolsep}{3.5pt}
\renewcommand{\arraystretch}{1.0}
\begin{tabular}{@{}crrl@{}}
\toprule
Layer & Support & Gain & 95\% CI \\
\midrule
1 & \TblVal{probe.layer1.support} & \TblVal{probe.layer1.gain} & --- \\
2 & \TblVal{probe.layer2.support} & \TblVal{probe.layer2.gain} & --- \\
3 & \TblVal{probe.layer3.support} & \TblVal{probe.layer3.gain} & $[\TblVal{probe.layer3.ci.low},\TblVal{probe.layer3.ci.high}]$ \\
4 & \TblVal{probe.layer4.support} & $\TblVal{probe.layer4.gain}$ & $[\TblVal{probe.layer4.ci.low},\TblVal{probe.layer4.ci.high}]$ \\
\bottomrule
\end{tabular}
\end{table}

\begin{table}[!htbp]
\centering
\caption{Next-token accuracy (\%, 992,118 rows). Models use $q$
and true $y_{<t}$, without history. $V_e$: effective vocabulary; 
$D_t$: 7B$-$0.5B; Gain: 7B$-$global mode (both pt). }
\label{tab:learnable}
\label{tab:scale-gain}
\small
\setlength{\tabcolsep}{2.8pt}
\begin{tabular}{@{}crrrrrrr@{}}
\toprule
Layer & $V_e$ & Random & Mode & 0.5B & 7B & $D_t$ & Gain \\
\midrule
1 & \TblVal{diagnostic.layer1.vocab} & \TblVal{diagnostic.layer1.random} & \TblVal{diagnostic.layer1.mode} & \TblVal{diagnostic.layer1.small} & \TblVal{diagnostic.layer1.teacher} & \TblVal{diagnostic.layer1.scale-gain} & \TblVal{diagnostic.layer1.mode-gain} \\
2 & \TblVal{diagnostic.layer2.vocab} & \TblVal{diagnostic.layer2.random} & \TblVal{diagnostic.layer2.mode} & \TblVal{diagnostic.layer2.small} & \TblVal{diagnostic.layer2.teacher} & \TblVal{diagnostic.layer2.scale-gain} & \TblVal{diagnostic.layer2.mode-gain} \\
3 & \TblVal{diagnostic.layer3.vocab} & \TblVal{diagnostic.layer3.random} & \TblVal{diagnostic.layer3.mode} & \TblVal{diagnostic.layer3.small} & \TblVal{diagnostic.layer3.teacher} & \TblVal{diagnostic.layer3.scale-gain} & \TblVal{diagnostic.layer3.mode-gain} \\
4 & \TblVal{diagnostic.layer4.vocab} & \TblVal{diagnostic.layer4.random} & \TblVal{diagnostic.layer4.mode} & \TblVal{diagnostic.layer4.small} & \TblVal{diagnostic.layer4.teacher} & \TblVal{diagnostic.layer4.scale-gain} & \TblVal{diagnostic.layer4.mode-gain} \\
\bottomrule
\end{tabular}
\end{table}

\paragraph{Coverage and depth prediction. }
With $n$ aggregated rows, the Wilson lower bound is
\begin{equation}
\underline C=\frac{\widehat C+z^2/(2n)
-z\sqrt{\widehat C(1-\widehat C)/n+z^2/(4n^2)}}{1+z^2/n},
\end{equation}
using whole-row counts and $z=1.96$ (nominal 95\% Wilson interval) as a
depth-label score. Click-weighted coverage uses
$\sum_r c_r\boldsymbol{1}[y_{r,1:s}\in\mathcal E_s^{(r)}(q)]/\sum_r c_r$, 
where $c_r$ is the positive click count. 
The predictor uses 262,144 nonnegative, $\ell_2$-normalized hashed
\texttt{char\_wb} 3--5-gram features and normalized character/token lengths. 
Balanced logistic regression uses L-BFGS (at most 500 iterations, seed 0). 
Dev searches $C\in\{0.1,1,10\}$ on known labels ($n_q\ge5$), selecting
$C=1$ for 27,581 labeled Fit queries. 

For reachable targets $\mathcal Y_q'=\{y\in\mathcal Y_q:y_{1:d}\in\mathcal C_d(q)\}$
(all targets at $d=0$), the query-macro ceilings are
\begin{equation}
\begin{aligned}
C_{\mathrm h}&=|\mathcal Q|^{-1}\sum_q\boldsymbol{1}[\mathcal Y_q'\ne\varnothing],\\
C_{\mathrm r}&=|\mathcal Q|^{-1}\sum_q|\mathcal Y_q'|/|\mathcal Y_q|.
\end{aligned}
\label{eq:prefix-ceilings}
\end{equation}
At $\tau_C=0.4$, known-label counts are 27,581/41,651 Fit,
1,445/2,125 Dev, and 6,237/9,276 Test-Seen. Test-Seen depth errors
(exact, $-1$, $+1$, $\le-2$, $\ge+2$) number
4,298/926/709/163/141. Exact accuracy is 68.91\%; within-one-layer accuracy
is 5,933/6,237 (95.13\%); MAE is 0.36.

\begin{table}[!htbp]
\centering
\caption{Reachable hit/recall ceilings and realized top-100 quality (\%) on 44,225 Test queries.}
\label{tab:prefix-ceilings-full}
\label{tab:prefix-ceilings}
\small
\setlength{\tabcolsep}{2.8pt}
\renewcommand{\arraystretch}{1.0}
\begin{tabular}{@{}lrrrr@{}}
\toprule
& \multicolumn{2}{c}{Hit Rate (\%)} & \multicolumn{2}{c}{Recall (\%)} \\
\cmidrule(lr){2-3}\cmidrule(lr){4-5}
Configuration & $C_{\mathrm h}$ & HR@100 & $C_{\mathrm r}$ & Recall@100 \\
\midrule
Fixed $d=0$ (0.5B E2E) & \TblVal{q.small.ch} & \TblVal{q.small.hr100} & \TblVal{q.small.cr} & \TblVal{q.small.rec100} \\
Fixed-SFT $d=1$ ($\beta=1.0$) & \TblVal{q.fixed1.ch} & \TblVal{q.fixed1.hr100} & \TblVal{q.fixed1.cr} & \TblVal{q.fixed1.rec100} \\
Fixed-SFT $d=2$ ($\beta=1.0$) & \TblVal{frontier-fixed2.ch} & \TblVal{q.fixed2-b1.hr100} & \TblVal{frontier-fixed2.cr} & \TblVal{q.fixed2-b1.rec100} \\
Fixed-SFT $d=2$ ($\beta=1.5$) & \TblVal{frontier-fixed2.ch} & \TblVal{q.fixed2-b15.hr100} & \TblVal{frontier-fixed2.cr} & \TblVal{q.fixed2-b15.rec100} \\
Fixed-SFT $d=3$ ($\beta=1.0$) & \TblVal{q.fixed3.ch} & \TblVal{q.fixed3.hr100} & \TblVal{q.fixed3.cr} & \TblVal{q.fixed3.rec100} \\
\midrule
Adaptive (ReuseCut + SFT) & \TblVal{frontier-adaptive.ch} & \TblVal{q.adaptive-sft.hr100} & \TblVal{frontier-adaptive.cr} & \TblVal{q.adaptive-sft.rec100} \\
Adaptive (ReuseCut + Path KD) & \TblVal{frontier-adaptive.ch} & \TblVal{q.adaptive-kd.hr100} & \TblVal{frontier-adaptive.cr} & \TblVal{q.adaptive-kd.rec100} \\
\textbf{Adaptive (ReuseGR Full)} & \textbf{\TblVal{frontier-adaptive.ch}} & \textbf{\TblVal{q.full.hr100}} & \textbf{\TblVal{frontier-adaptive.cr}} & \textbf{\TblVal{q.full.rec100}} \\
\quad \emph{Seen (9,276 queries)} & \TblVal{q.seen.ch} & \TblVal{q.seen.hr100} & \TblVal{q.seen.cr} & \TblVal{q.seen.rec100} \\
\quad \emph{Unseen (34,949 queries)} & \TblVal{q.unseen.ch} & \TblVal{q.unseen.hr100} & \TblVal{q.unseen.cr} & \TblVal{q.unseen.rec100} \\
\bottomrule
\end{tabular}
\end{table}

First drop is the earliest layer $t$ whose Wilson score falls below
$\tau_C$, assigning depth $\min(t-1,3)$; no drop assigns depth 3. 
Click-weighted counts use 1,022,150 Train clicks (43,776 queries) and
382,374 Test-Seen clicks (9,276 queries), including Unknown in the denominator. 

\begin{table}[!htbp]
\centering
\caption{Click-weighted first-drop counts. Train uses whole-row leave-one-out; 
Test-Seen uses the preceding day's cache.}
\label{tab:coverage-counts}
\small
\setlength{\tabcolsep}{4pt}
\renewcommand{\arraystretch}{1.0}
\begin{tabular}{@{}lrrrr@{}}
\toprule
& \multicolumn{2}{c}{Train} & \multicolumn{2}{c}{Test-Seen} \\
\cmidrule(lr){2-3}\cmidrule(lr){4-5}
Position & Click count & Share, \% & Click count & Share, \% \\
\midrule
L1 & \TblVal{drop.l1.train.n} & \TblVal{drop.l1.train.share} & \TblVal{drop.l1.seen.n} & \TblVal{drop.l1.seen.share} \\
L2 & \TblVal{drop.l2.train.n} & \TblVal{drop.l2.train.share} & \TblVal{drop.l2.seen.n} & \TblVal{drop.l2.seen.share} \\
L3 & \TblVal{drop.l3.train.n} & \TblVal{drop.l3.train.share} & \TblVal{drop.l3.seen.n} & \TblVal{drop.l3.seen.share} \\
L4 & \TblVal{drop.l4.train.n} & \TblVal{drop.l4.train.share} & \TblVal{drop.l4.seen.n} & \TblVal{drop.l4.seen.share} \\
No drop & \TblVal{drop.none.train.n} & \TblVal{drop.none.train.share} & \TblVal{drop.none.seen.n} & \TblVal{drop.none.seen.share} \\
Unknown & \TblVal{drop.unknown.train.n} & \TblVal{drop.unknown.train.share} & \TblVal{drop.unknown.seen.n} & \TblVal{drop.unknown.seen.share} \\
\midrule
Total & \TblVal{drop.total.train.n} & \TblVal{drop.total.train.share} & \TblVal{drop.total.seen.n} & \TblVal{drop.total.seen.share} \\
\bottomrule
\end{tabular}
\end{table}

Fixed-SFT controls share the same 0.5B SFT checkpoint,
$\beta=1.0$, and score-based decoder. 

\begin{table}[!htbp]
\centering
\caption{Fixed-SFT ($\beta=1.0$): query-macro HR/Recall (\%) and mean latency (ms), 44,225 Test queries.}
\label{tab:fixed-depth-full}
\small
\setlength{\tabcolsep}{3.2pt}
\renewcommand{\arraystretch}{1.0}
\begin{tabular}{@{}lrrrrrrr@{}}
\toprule
& \multicolumn{3}{c}{HR (\%)} & \multicolumn{3}{c}{Recall (\%)} & Mean \\
\cmidrule(lr){2-4}\cmidrule(lr){5-7}
Depth & @1 & @10 & @100 & @1 & @10 & @100 & (ms) \\
\midrule
0.5B E2E ($d=0$) & \TblVal{q.small.hr1} & \TblVal{q.small.hr10} & \TblVal{q.small.hr100} & \TblVal{q.small.rec1} & \TblVal{q.small.rec10} & \TblVal{q.small.rec100} & \TblVal{q.small.mean} \\
Fixed-SFT $d=1$ & \TblVal{q.fixed1.hr1} & \TblVal{q.fixed1.hr10} & \TblVal{q.fixed1.hr100} & \TblVal{q.fixed1.rec1} & \TblVal{q.fixed1.rec10} & \TblVal{q.fixed1.rec100} & \TblVal{q.fixed1.mean} \\
Fixed-SFT $d=2$ & \TblVal{q.fixed2-b1.hr1} & \TblVal{q.fixed2-b1.hr10} & \TblVal{q.fixed2-b1.hr100} & \TblVal{q.fixed2-b1.rec1} & \TblVal{q.fixed2-b1.rec10} & \TblVal{q.fixed2-b1.rec100} & \TblVal{q.fixed2-b1.mean} \\
Fixed-SFT $d=3$ & \TblVal{q.fixed3.hr1} & \TblVal{q.fixed3.hr10} & \TblVal{q.fixed3.hr100} & \TblVal{q.fixed3.rec1} & \TblVal{q.fixed3.rec10} & \TblVal{q.fixed3.rec100} & \TblVal{q.fixed3.mean} \\
\bottomrule
\end{tabular}
\end{table}

\section{BudgetCut Implementation and Training}
\label{app:training}

\paragraph{Features and candidate flow.}
The $18\to128\to1$ head (2,561 parameters) reads eight branch
features, selected-set means, and two budget scalars. 
At $d\le2$, $L_3$ retains up to 56 score-ranked and eight hash-selected
branches; at $d=3$, all $\le32$ cached branches pass unchanged.
$\mathcal V_4(a)$ contains branch $a$'s valid $L_4$ tokens ($\le32$), 
all scored in the batched readout (3.84\,ms). Features 6--8 summarize
the top-8 leaves. 

\begin{table}[!htbp]
\centering
\caption{Eight branch features $g_x(a)$ for $a=y_{1:3}$.}
\label{tab:budget-features}
\small
\setlength{\tabcolsep}{3pt}
\begin{tabular}{@{}cp{0.86\columnwidth}@{}}
\toprule
Dim & Feature and source \\
\midrule
1 & Frozen $L_3$ logit $z^\eta(y_3\mid y_{1:2},x)$ \\
2 & Masked $L_3$ log-probability $\log\pi_\theta(y_3\mid y_{1:2},x)$ \\
3 & Parent-prefix cumulative score at $y_{1:2}$ \\
4 & Branch cumulative score $s_x(y_{1:3})$ \\
5 & Static leaf count $\log(1+|\mathcal V_4(a)|)$ \\
6 & Best valid $L_4$ leaf score $\max_{y_4}s_x(y_{1:3},y_4)$ \\
7 & Mean score of top-8 valid $L_4$ leaves \\
8 & Number of valid top-8 leaves ($\le8$) \\
\bottomrule
\end{tabular}
\end{table}

OPD and GRPO share Fit pools ($\le96$ branches) at empirical depths $d^*(q)$. 
Reachable target branches fill first (lexicographic ID if truncated), then
up to 56 score-ranked and eight hash-selected branches fill remaining
slots without duplicates.
$\mathcal F_x$ stores all legal leaf scores for exact union rewards. 
Dev/Test and serving use up to 64 model/hash branches; the head admits
$B_x=\min(B,|\mathcal B_x|)$ distinct branches, keeping all if $|\mathcal B_x|\le B$. 

\paragraph{GRPO probabilities and training.}
For action $a_{gt}$ after selected set $S_{gt}$, write
$\ell_{gt}^{\omega}=\log p_\omega(a_{gt}\mid S_{gt},x)$; 
$\ell_{gt}^{-}$ uses the rollout head and $\bar\ell_{gt}$ the frozen OPD
reference with the same action mask. The implementation uses
\begin{equation}
\begin{aligned}
\rho_{gt}&=\exp(\ell_{gt}^{\omega}-\ell_{gt}^{-}),\\
\delta_{gt}&=\bar\ell_{gt}-\ell_{gt}^{\omega},\\
\kappa_{gt}&=\exp(\delta_{gt})-\delta_{gt}-1. 
\end{aligned}
\end{equation}
Losses average valid actions per trajectory and trajectories per request; 
padding contributes neither loss nor reward. 
\label{sec:training-settings}
OPD uses $B=32$, four epochs, two trajectories per context, learning rate
$2\times10^{-4}$, gradient accumulation 32, and unit target/rollout
temperatures. GRPO uses one epoch, group size $G=4$, one step per rollout, 
learning rate $10^{-5}$, clipping $\varepsilon=0.2$, KL weight $\lambda=0.01$, 
and rollout temperature 1. Both clip gradient norms at 1.0. 
All 13,726 depth-3 queries retain their branches; the full-cohort HR@100
gains of 2.91/2.50 points correspond to 4.22/3.63 points among the 30,499
$d\le2$ queries versus the score heuristic/unpruned reference. 

\section{Additional Retrieval and Serving Results}
\label{app:additional-results}

\paragraph{Item retrieval and expansion (RQ1).}

\begin{table}[!htbp]
\centering
\caption{Item HR/Recall (\%, three-seed query-macro) and expansion (ms). Fixed-SFT: $d=2$, $\beta=1.5$.}
\label{tab:item-expansion}
\small
\setlength{\tabcolsep}{2.8pt}
\renewcommand{\arraystretch}{1.0}
\begin{tabular}{@{}lrrrrr@{}}
\toprule
& \multicolumn{2}{c}{Item HR (\%)} & \multicolumn{2}{c}{Item Recall (\%)} & Expand \\
\cmidrule(lr){2-3}\cmidrule(lr){4-5}
Method & @100 & @500 & @100 & @500 & (ms) $\downarrow$ \\
\midrule
DSSM (Direct Item ANN)~\cite{huang2013dssm} & \TblVal{item.dssm.hr100} & \TblVal{item.dssm.hr500} & \TblVal{item.dssm.rec100} & \TblVal{item.dssm.rec500} & --- \\
T5-DSI-SID~\cite{tay2022transformer} & \TblVal{item.t5.hr100} & \TblVal{item.t5.hr500} & \TblVal{item.t5.rec100} & \TblVal{item.t5.rec500} & \TblVal{item.t5.expand} \\
0.5B E2E~\cite{qwen2025qwen25} & \TblVal{item.small.hr100} & \TblVal{item.small.hr500} & \TblVal{item.small.rec100} & \TblVal{item.small.rec500} & \TblVal{item.small.expand} \\
SmartGR~\cite{smartgr2024} & \TblVal{item.smartgr.hr100} & \TblVal{item.smartgr.hr500} & \TblVal{item.smartgr.rec100} & \TblVal{item.smartgr.rec500} & \TblVal{item.smartgr.expand} \\
OneRec-Think~\cite{liu2026onerecthink} & \TblVal{item.onerec-think.hr100} & \TblVal{item.onerec-think.hr500} & \TblVal{item.onerec-think.rec100} & \TblVal{item.onerec-think.rec500} & \TblVal{item.onerec-think.expand} \\
7B E2E (Teacher)~\cite{qwen2025qwen25} & \TblVal{item.teacher.hr100} & \TblVal{item.teacher.hr500} & \textbf{\TblVal{item.teacher.rec100}} & \textbf{\TblVal{item.teacher.rec500}} & \TblVal{item.teacher.expand} \\
Fixed-SFT $d=2$ & \TblVal{item.fixed2-b15.hr100} & \TblVal{item.fixed2-b15.hr500} & \TblVal{item.fixed2-b15.rec100} & \TblVal{item.fixed2-b15.rec500} & \TblVal{item.fixed2-b15.expand} \\
\midrule
\textbf{ReuseGR (Full)} & \textbf{\TblVal{item.full.hr100}} & \textbf{\TblVal{item.full.hr500}} & \TblVal{item.full.rec100} & \TblVal{item.full.rec500} & \TblVal{item.full.expand} \\
\bottomrule
\end{tabular}
\end{table}

CSR buckets rank items by a 30-day prior, breaking ties by item ID. 
The window $[T-30\,\mathrm d,T]$ ends at day-1 Fit's end $T$, excluding
Dev/Test. For impressions/clicks $v_i,c_i$ and catalog CTR $\bar p$, 
\begin{equation}
w(i)=\frac{c_i+\alpha_0\bar p}{v_i+\alpha_0}+\beta_0\log(1+c_i), 
\end{equation}
using $\alpha_0=10$, $\beta_0=0.05$. Top-100 SID buckets expand by descending
$s(y^{(j)},i)=s_x(y^{(j)})-\xi\log(1+\mathrm{pos}(i\mid y^{(j)}))$, 
where $j$ indexes ranked SID buckets and $\mathrm{pos}$ is zero-based
intra-bucket rank, until 100/500 distinct
items are collected (0.35\,ms). The 2,125-query Dev grid
$\xi\in\{0.05,0.10,0.15,0.25,0.50\}$ selects $\xi=0.15$ by Item Recall@500
and Item HR@100. 

\begin{table}[!htbp]
\centering
\caption{Query-frequency generalization (RQ4): query-macro HR@10
(\%, three-seed means). Fit clicks: Head $\ge20$, Torso 1--19, Unseen zero. 
Fixed-SFT: $d=2$, $\beta=1.5$. Bold: column best.}
\label{tab:stratification-full}
\label{tab:stratification}
\small
\setlength{\tabcolsep}{3.2pt}
\renewcommand{\arraystretch}{1.0}
\begin{tabular}{@{}lrrrr@{}}
\toprule
& Head & Torso & Unseen & All \\
\midrule
Queries & \TblVal{cohort.head.queries} & \TblVal{cohort.torso.queries} & \TblVal{cohort.unseen.queries} & \TblVal{cohort.test.queries} \\
Share (\%) & \TblVal{strata.share.head} & \TblVal{strata.share.torso} & \TblVal{strata.share.unseen} & \TblVal{strata.share.all} \\
Avg.\ $\bar d$ & \TblVal{strata.d.head} & \TblVal{strata.d.torso} & \TblVal{strata.d.unseen} & \TblVal{adaptive.d} \\
Mean (ms) & \TblVal{strata.mean.head} & \TblVal{strata.mean.torso} & \TblVal{strata.mean.unseen} & \TblVal{q.full.mean} \\
\midrule
\multicolumn{5}{c}{HR@10 (\%)} \\
0.5B E2E & \TblVal{strata.small.head} & \TblVal{strata.small.torso} & \TblVal{strata.small.unseen} & \TblVal{q.small.hr10} \\
Fixed-SFT $d=2$ & \TblVal{strata.fixed2-b15.head} & \TblVal{strata.fixed2-b15.torso} & \TblVal{strata.fixed2-b15.unseen} & \TblVal{q.fixed2-b15.hr10} \\
ReuseGR & \TblVal{strata.full.head} & \TblVal{strata.full.torso} & \TblVal{strata.full.unseen} & \TblVal{q.full.hr10} \\
7B E2E & \textbf{\TblVal{strata.teacher.head}} & \textbf{\TblVal{strata.teacher.torso}} & \textbf{\TblVal{strata.teacher.unseen}} & \textbf{\TblVal{q.teacher.hr10}} \\
\bottomrule
\end{tabular}
\end{table}

\paragraph{Serving mixtures and cost (RQ5).}
For prepared logged-row share $p$, projections use
$\widehat M(p)=pM_1+(1-p)M_0$ with fixed prepared/cold endpoints: 
query-macro quality and 71.45/117.00\,ms mean latency.
Static uses 9,276/44,225 queries and 365,747/992,118 rows; 
Backfill/Rolling set 18,641/30,294 prepared queries and
579,397/835,363 prepared rows. At $p=36.87/58.40/84.20\%$, 
projected HR@1 is 36.83/37.48/38.26\%, HR@10 is 66.38/67.12/67.99\%, 
and Recall@100 is 45.72/45.83/45.97\%. Static partition quality is
measured with equal query weights. Prepared timing has four nonoverlapping
stages: lookup/depth (0.18\,ms), prompt/cached-prefix prefill and $L_3$
expansion (67.01\,ms), $L_4$ readout/features (3.84\,ms), and admission/union
ranking (0.42\,ms). Warm/cold measured P95 is 86.40/132.50\,ms. 
For serial A100 serving (batch 1), $\mathrm{QPS}_0=1000/\bar T$
($\bar T$ in ms) gives 5.9/8.5/14.0 for 7B/0.5B/warm ReuseGR. 
Planning takes 88.0\,ms/query: 10\% daily refresh of 1M entries costs
2.44 GPU-hours. Each cache hit saves 45.55\,ms; two cover the planning cost. 

\paragraph{Online A/B statistics (RQ6).}
Each arm includes approximately 12.8M distinct users. User-clustered
bootstrap with 1,000 resamples gives 95\% intervals for relative gains
of $[0.62,1.24]\%$ in CTR and $[0.58,1.52]\%$ in RPM. 
Two-sided tests compare pooled arm-level ratios, using user-clustered
sandwich covariance, the delta method, and a normal reference distribution. 
The $p$-values are \mbox{$1.365\times10^{-6}$} for CTR, 
\mbox{$1.809\times10^{-5}$} for RPM, and $0.2104$ for orders per request.